\documentclass[aps,prx,twocolumn,groupedaddress]{revtex4-2}

\usepackage{amsmath}
\usepackage{amssymb}
\usepackage{mathtools}
\usepackage{physics}
\usepackage[noabbrev]{cleveref}
\usepackage{float}

\newcommand{\N}{\mathbb{N}}	
\newcommand{\vqd}{v_{QD}}

\begin{document}

    \title{Non-adiabatic quantum control of valley states in silicon}


    \author{Alan Gardin$^{1}$}
    \email{alan.gardin@adelaide.edu.au}
    \author{Ross D. Monaghan$^1$}
    \author{Tyler Whittaker$^1$}
    \author{Rajib Rahman$^2$}
    \author{Giuseppe C. Tettamanzi$^1$}
    \email{giuseppe.tettamanzi@adelaide.edu.au}
    \affiliation{$^1$School of Physics \& Institute of Photonics and Advanced Sensing, The University of Adelaide, Adelaide SA 5005, Australia.}
    \affiliation{$^2$School of Physics, The University of New South Wales, Sydney, NSW 2052, Australia.}


    \date{\today}
    \begin{abstract}
   	    Non-adiabatic quantum effects, often experimentally observed in semiconductors nano-devices such as single-electron pumps operating at high frequencies, can result in undesirable and uncontrollable behaviour.
        However, when combined with the valley degree of freedom inherent to silicon, these unfavourable effects may be leveraged for quantum information processing schemes.
        By using an explicit time evolution of the Schrodinger equation, we study numerically non-adiabatic transitions between the two lowest valley states of an electron in a quantum dot formed in a SiGe/Si heterostructure.
        The presence of a single atomic layer step at the top SiGe/Si interface opens an anti-crossing in the electronic spectrum as the centre of the quantum dot is varied.
        We show that an electric field applied perpendicularly to the interface allows tuning of the anti-crossing energy gap.
        As a result, by moving the electron through this anti-crossing, and by electrically varying the energy gap, it is possible to electrically control the probabilities of the two lowest valley states.
    \end{abstract}


    \maketitle

    \graphicspath{{.}}

    \section{Introduction}
    \textbf{Adiabatic theorem and manifestations in single-electron pumps.}
 	According to the adiabatic theorem, a system initially in its ground state, and whose Hamiltonian evolves slowly in time, is expected to remain in the ground state of the instantaneous Hamiltonian, provided the ground state is non-degenerate.
	However, if the Hamiltonian varies quickly, then this approximation is no longer valid and the state is best described by a superposition of eigenstates of the instantaneous Hamiltonian.
	This superposition of states induces spatial oscillations of the wave-function, which are an observable manifestation of the violation of the adiabatic theorem.
	This is encountered in single electron pumps operating at fast pumping frequencies, where being in the non-adiabatic regime leads to a decrease in the accuracy and precision of the pump \cite{2015KaestnerKashcheyevs}.
	For instance, in the work of \citeauthor{2019Yamahata} \cite{2019Yamahata} non-adiabatic oscillations were experimentally measured in single-electron pumps formed by a silicon nanowire field effect transistor.
	They also presented a one-dimensional simulation of non-adiabatic oscillations between electron orbitals for a moving quantum dot and demonstrated the resulting spatial oscillations of the wave function.

    \textbf{Valleys in silicon and SiGe structures. }
    In a [001] grown Si$_{x}$Ge$_{1-x}$ heterostructure, the tensile strain between the Si and SiGe layers breaks the six-fold valley degeneracy of bulk silicon, leading to a four-fold degeneracy ($k_x$ and $k_y$ valleys) raised in energy and a two-fold degeneracy ($k_z$ valleys) lowered in energy.
    The latter can be further broken by interface roughness, impurities, or electric fields \cite{2013Zwanenburg,rahman2011}.
    It is sufficient to focus on the two lowest $k_z$ valley states because of the high energy gap (typically around 200 meV \cite{2013Zwanenburg}) between them and the other $k_x$ and $k_y$ valley states.

    \textbf{Non-adiabatic transitions with valleys is unexplored.}
    Whilst \citeauthor{2019Yamahata} considered electron orbitals, they do not take into account the valley degrees of freedom inherent to silicon, which are expected to couple with the orbital states in any realistic device \cite{2010FriesenCoppersmith, 2013Zwanenburg, 2019TariqHu}.
	Hence it may be possible, due to valley-orbit coupling, to observe non-adiabatic transitions between valley states.
	Note that coherent oscillations between valley states has been previously studied in pure valley qubit schemes based on double quantum dots~\cite{2017SchoenfieldFreemanJiang,2019Penthorn}, while we instead focus on a single quantum dot.
	Similarly, \citeauthor{2016Boross} \cite{2016Boross} considered coherent oscillations between valley states in a single dot, but their driving scheme is limited to a weak electric field, contrary to our scheme.

    \textbf{Model and contribution of the paper.}
    In this paper, we study the dynamics of a single electron trapped in a quantum dot formed by a [001] grown Si$_{0.8}$Ge$_{0.2}$/Si heterostructure.
    The appearance of valley physics stems from our modelling of the SiGe/Si heterostructure by use of a two-band model.
    Non-adiabatic effects are then introduced by quickly varying the centre of the quantum dot potential.
    We then successively consider two two-dimensional models of a simplified SiGe/Si heterostructure, both introduced in section \ref{sec:models}.

    In the first model, we present an ideally grown heterostructure, where the orbital degrees of freedom originating from the quantum dot are decoupled from the valley degrees of freedom, and thusly observe only non-adiabatic transitions between orbital states.
    An analytic study of such transitions is tackled in section \ref{sec:non-adiabatic-ideal}.

    In the second model, we model a miscut SiGe/Si heterostructure by introducing a single atomic step at the interface, akin to that of \citeauthor{2016Boross}.
    In section \ref{sec:spectrum}, we explain why the introduction of this step alters the spectrum of the system as the quantum dot position is varied.
    The key physics is that an anti-crossing between the two lowest valley states opens, with an energy gap tuneable by an electric field applied along the $z$ direction.
    This behaviour is verified by a realistic tight-binding calculation using NEMO 3D's 20 band $sp^{3}d^{5}s^{*}$ model ~\cite{2002Oyafuso}. \
    We detail this in appendix \ref{app:validation}.
    As a result of the anti-crossing, the non-adiabatic pulsing of the electron through this step allows one to induce transitions from one valley state to another.
    Section \ref{sec:non-adiabatic-step} discusses the main result of the paper, namely that the final state can be electrically controlled by varying the anti-crossing energy gap.
    Due to the appearance of the avoided crossing in the spectrum, we model the transition using the Landau-Zenner approximation ~\cite{1981Rubbmark} and achieve good concordance.
    From studying the results there exists a range of driving parameters in which the final state forms a good two-level system, opening the door for qubit applications.
    We discuss the viability of such non-adiabatic valley qubit in section \ref{sec:qubit}.

    \section{\label{sec:models}Models}
    \subsection{\label{subsec:ideal-model}Ideal model}
\begin{figure}[h]
        \centering
        \includegraphics[width=\columnwidth]{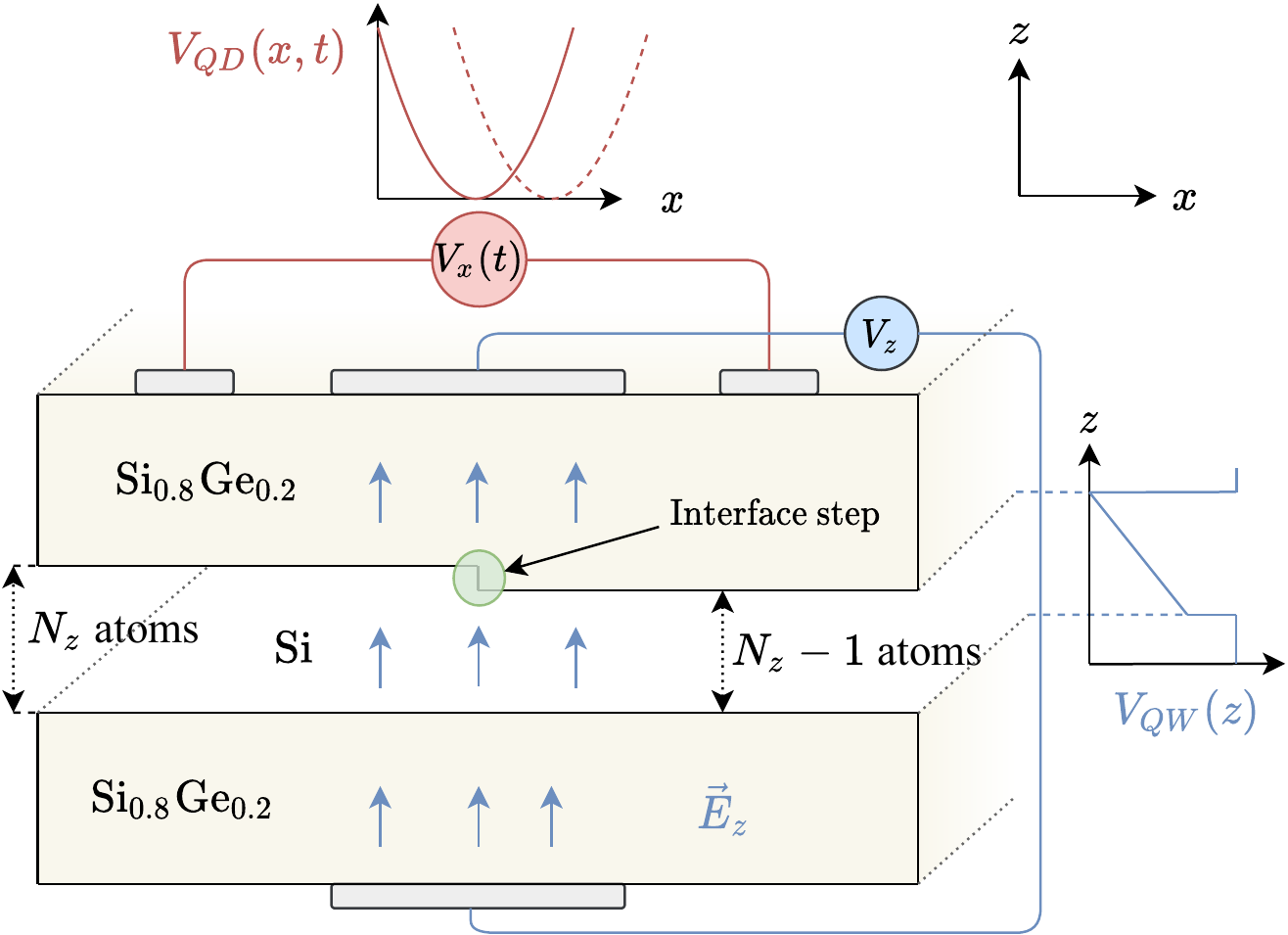}
        \caption{Geometry of the two-dimensional SiGe heterostructure model with the $y$ axis neglected. We apply a time-dependent harmonic quantum dot potential $V_{QD}$ in $x$, and an electric field in $z$. The quantum well in the $z$ axis, originating from the SiGe/Si interfaces, is composed of $N_z$ atoms and decreases by one atomic layer at the step (green circle). The dashed parabola represents the potential profile at $t' > t$, reflecting the spatial movement of the harmonic potential.}
        \label{fig:2d-heterostructure}
    \end{figure}

    \textbf{Quantum dot potential.}
    We now present the two-dimensional model, in the $(x,z)$ plane, used throughout this paper.
    We assume the growth axis $z$ of our heterostructure to be orthogonal to the $x$ axis, as shown in \cref{fig:2d-heterostructure}.
    In the $x$ direction, the electron has a transverse effective mass $m_x = 0.19 \, m_e$.
    The quantum dot potential $V_{QD}$ is a harmonic potential of angular frequency $\omega$ whose minimum follows a trajectory $\xi$, hence $V_{QD}(x,t) = \frac{1}{2} m_x \omega^2(x-\xi(t))^2$.
    We choose $\omega$ such that we obtain an orbital level spacing $\Delta \epsilon = \hbar \omega =$ 2 meV for the orbital states.
    This value is chosen to isolate the higher orbital states from the two lowest valley states.
    Note that the $y$ axis is neglected in our work since it does not affect the physics discussed.

    \textbf{Quantum dot trajectory.}
    The evolution in time of the quantum dot position is the only source of time-dependence in Hamiltonian.
    The quantum dot is initially at position $x_0=0$ nm at $t_0=0$ ps, and moves with constant velocity $\vqd$ until it reaches $x_1$ at time $t_1$.
    The parameters $x_1$ and $t_1$ are varied to obtain different values for $\vqd$.
    The trajectory $\xi$ of the quantum dot can be written:
    \begin{equation}
        \xi(t) = \begin{cases*}
            x_0  \text{ for } t \leqslant t_0\\
            x_0 +v_{QD}(t-t_0) \text{ for } t_0 \leqslant t \leqslant t_1\\
            x_1 \quad \quad \;\, \text{ for } t_1 \leqslant t
        \end{cases*}
    \label{eq:linear-trajectory}
    \end{equation}
    The Hamiltonian describing the dynamics along the $x$ axis is
    \begin{equation}
        H_{x}(x,t) = \frac{p_x^2}{2m_x} + V_{QD}(x,t)
        \label{eq:hamiltonian-x}
    \end{equation}
    and is identical for all $z$ in the ideal model.

    \textbf{Description of the $z$ confinement. }
    The $z$ confinement and the valley effects are modelled using the tight-binding model for strained silicon quantum wells of \citeauthor{2004BoykinPRB} \cite{2004BoykinAPL,2004BoykinPRB}.
    The resulting Hamiltonian for the $z$ slices depends on three parameters: the onsite energy $\eta$, the nearest neighbor hopping $v$ and the second-neighbor hopping $u$ giving the Hamiltonian in \cref{eq:boykin-hamiltonian}.
    \begin{equation}
        H_z(\epsilon, u, v) =
        \begin{pmatrix}
            \eta & v & u & 0 & \dots & \dots & 0\\
            v & \eta & v & u & 0 & \ddots & \vdots \\
            u & v & \eta & v & u & \ddots & \vdots \\
            0 & \ddots & \ddots & \ddots & \ddots & \ddots & 0 \\
            \vdots & \ddots & u & v & \eta & v & u \\
            \vdots & \ddots & 0 & u & v & \eta & v \\
            0 & \dots & \dots & 0 & u & v & \eta  \\
        \end{pmatrix}
        \label{eq:boykin-hamiltonian}
    \end{equation}
    In this Hamiltonian, each site represents an atom of the quantum well in the $z$ axis.
    We will define $N_z=40$ to be the total number of atoms in the $z$ axis, where each atom is separated by $1.366$\AA.
    We will also refer to $N_z$ as the quantum well width, expressed in number of atoms.
    A monolayer corresponds to two atoms, so that our choice of quantum well width corresponds to 20 monolayers.
    Note that varying the width of the Si/SiGe quantum well, as we could do in the simulations, is experimentally feasible.

    \textbf{Expected eigenstates and spectrum. }
    In our model, the quantum well width is $L_z = N_z \times 0.1366 = 5.464$ nm.
    The energies of this infinite quantum well barrier are given by $E_m= m^2\frac{\hbar^2 \pi^2}{2m_l L_z^2}$ with $m_l = 0.92 m_e$ the electron's longitudinal effective mass and $m \in \mathbb N^*$ the orbital number.
    The energy gap between the first and second quantum well eigenstates is then $3\frac{\hbar^2 \pi^2}{2m_l L_z^2}=13$ meV.
    Because of this high energy gap, it is reasonable to only consider the first eigenstate of the quantum well.
    The eigenstates of $H_x$, which are eigenstates of the harmonic oscillator, are denoted by $\varphi_n$ with $n\in \mathbb N$ describing the orbital number.

    The Hamiltonian describing the $z$ confinement gives a valley splitting less than the orbital spacing $\Delta \epsilon = \hbar \omega$ induced by to the harmonic oscillator.
    As a consequence, we expect to observe two well defined valley states for each orbital.
    We will denote the corresponding eigenstates $\psi_{n,v}$ and energies $\epsilon_{n,v}$, with $(n,v)$ where $n\in \mathbb{N}$ is the orbital index and $v\in\qty{-,+}$ the valley index.
    It will be convenient later on to index the states and the energies with a single index for readability.
    Since the $v=-$ valley states always have lower energies than the $v=+$ states for a given $n$, we will use the natural mapping $(0,-)\to 0$, $(0,+)\to 1$, $(1,-)\to 2$ and so on to label the energies and eigenstates by increasing energies.
    The meaning of the indices will be understood from context depending on whether there is one or two indices.

    As a final note, the full Hamiltonian $H$ for this idealised model is separable in $x$ and $z$, and hence the valley degrees of freedom do not couple to the orbital degree of freedom.
    The eigenstates are the product of the eigenstates of the Hamiltonians $H_x$ and $H_z$, and indeed the indices $(n,v)$ are good quantum numbers.

    \subsection{\label{subsec:step-model}Single step model}
    \textbf{Description of the interface step, and experimental justification.}
    As in the work of ~\citeauthor{2016Boross}, we introduce an interface step consisting of a single atomic layer at $x_s=30$ nm, modelled by decreasing the width of the quantum well for all $z$ slices where $x \geqslant x_s$ (see \cref{fig:2d-heterostructure}).
    This interface step is motivated by the single atomic layer steps observed in SiGe/Si heterostructures grown on slightly vicinal [0 0 1] Si substrates \cite{2002-teichert,2002-myslivecek,2012-evans,1994-phang}.
    In particular, these steps occur at polar miscut angles~$\theta \leq 2^{\circ}$, with terrace width on the order of 10nm \cite{2002-myslivecek,2012-evans,1994-phang}.
    Given $\theta$ can be controlled by manufacturers to within $0.1^\circ$ \cite{2002-teichert} and uniform terrace width can be achieved as in \cite{2002-myslivecek}, one could manufacture a structure with an interface step similar to that modelled in this paper.

    \textbf{Differences with~\citeauthor{2016Boross}.}
    Unlike \citeauthor{2016Boross}, we do not allow the wave function to penetrate inside the SiGe layer, instead enforcing hard-wall boundary conditions, however, the physics observed is expected to be similar.
    And indeed as shown in \cref{fig:step-spectra}, the evolution of the spectrum is similar to that obtained in \citeauthor{2016Boross}.
    Another difference in our work is that we do not limit ourselves to weak perturbations along the $x$ axis: we study the opposite limit of strong non-adiabatic perturbations.
    Furthermore, our modelling has been validated with NEMO 3D's 20 band $sp^{3}d^{5}s^{*}$ model, see appendix \ref{app:validation}.

    \subsection{Methods}
    We numerically solve the ground state of the total Hamiltonian $H$ at the initial time $t_0=0$ ps and use this result as the initial wave function for solving the time-dependent Schrödinger equation.
    The time-dependent Schrödinger equation is solved using the Crank-Nicolson scheme \cite{1947CrankNicolson}.
    The onsite energy parameter $\eta$ of $H_z$ (\cref{eq:boykin-hamiltonian}) is set to 1.395 eV such that the spectrum has negative values close to zero, allowing for easier differentiation between ``real'' eigenvectors and ``artificial'' eigenvectors (eigenvalues of 0).
    Indeed, this is necessitated by the setting the column and row of the relevant lattice sites to zero in the Hamiltonian to produce the step in the interface, described above, as this introduces null eigenvalues.
    The difference between our chosen value of $\eta$ and the choice made in \citeauthor{2004BoykinPRB} is then restored after diagonalisation.
    The values used for the off-diagonal elements can be found in figure 2 of \citeauthor{2004BoykinPRB} \cite{2004BoykinPRB}.

    In all simulations, we fix $x_0=0$ nm and $t_0=0$ ps.
    For the simulations involving the ideal model, we fix $t_1-t_0$ to 5 ps and vary $x_1$ to change $\vqd$.
    This choice ensures that the perturbation is applied for the same amount of time in all simulations.
    For the step model however, we fix $x_1 = 55$ nm and vary $t_1$ to change $\vqd$.
    As discussed later, the spatial evolution of the dot relative to the step is critical for analysis of the simulations, such that fixing the initial and final positions and changing the time is the more appropriate choice of altering the velocity.

    \section{\label{sec:non-adiabatic-ideal}Non adiabatic effects in the ideal model}
    \subsection{Preliminary remarks}
\begin{figure}[h]
        \centering
        \includegraphics[width=\columnwidth]{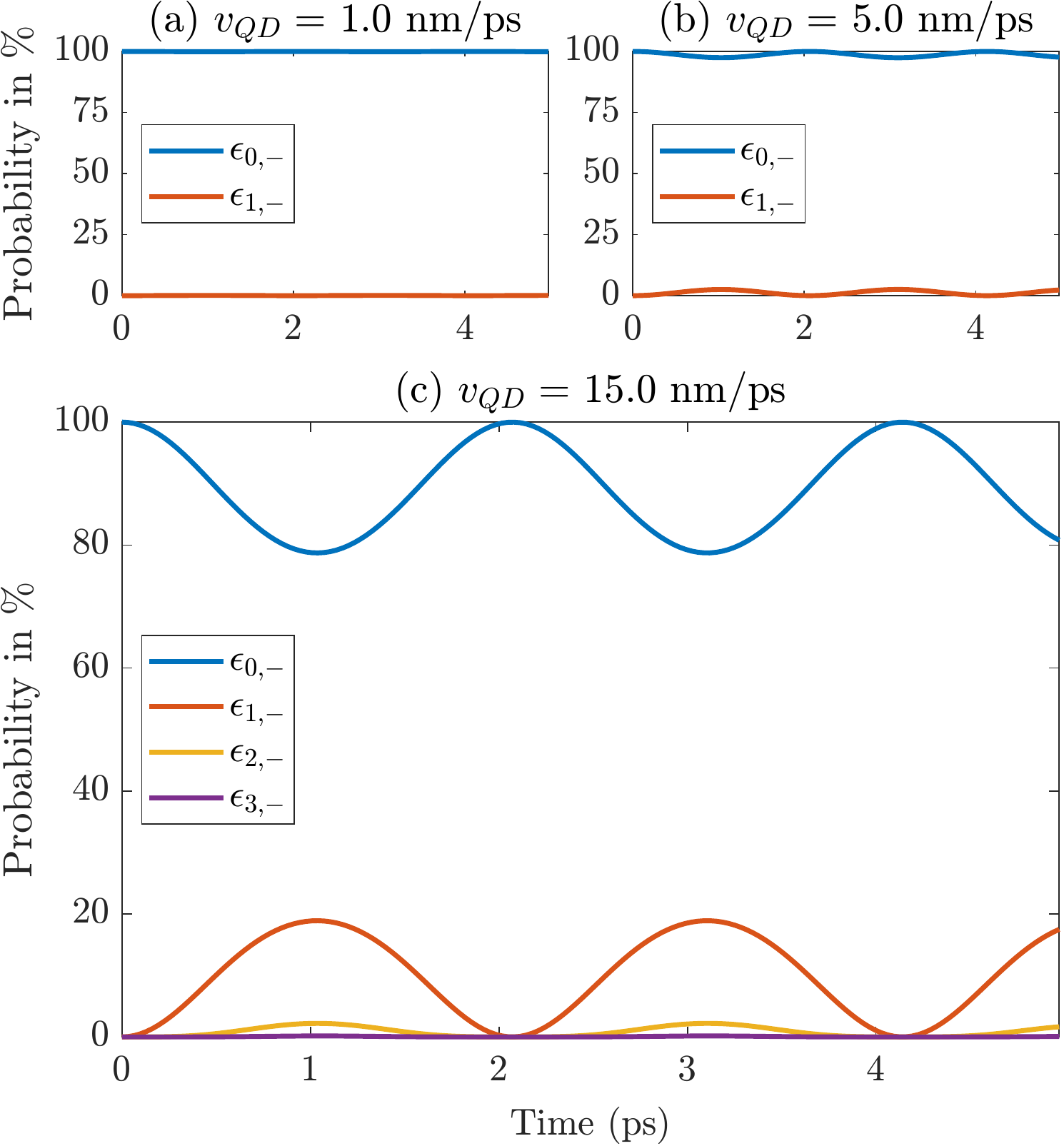}
        \caption{
            Time evolution of the probabilities for the ideal model (no step and hence no valley physics).
            For small speeds, the system mostly occupies the two lowest energy orbitals, forming a two-dimensional Hilbert space, as needed for a qubit.
            However, a higher speed introduces higher excited states and does not behave as a two-level system anymore.
            Note that the period of oscillations is constant and the higher the speed the bigger the amplitudes of the probability oscillations.}
        \label{fig:ideal-probs}
    \end{figure}

    \textbf{Introduction.}
    In this section, we comment on the numerical results obtained with the ideal model (\cref{fig:ideal-probs}) of section \ref{subsec:ideal-model}, i.e when no interface step is present.
    As explained in the previous section, the Hamiltonian is separable in the $x$ and $z$ coordinates.
    Since the time-dependence only originates from the $x$ axis, the dynamics of the system reduce to that of a driven quantum harmonic oscillator described by the Hamiltonian $H_x$ of \cref{eq:hamiltonian-x}.
    As a consequence, the only parameter we vary in our simulations is the quantum dot speed $\vqd$ (in particular, the applied electric field $E_z$ does not influence the numerical results).
    This model is essentially the same as the one from \citeauthor{2019Yamahata} \cite{2019Yamahata}, with the difference being the trajectory $\xi$ of the quantum dot.
    Our choice for the trajectory $\xi$ (see \cref{eq:linear-trajectory}) will simplify the interpretation of the numerical results.

    \textbf{Oscillatory features of the probabilities: amplitude and period. }
    As a preliminary remark, one can notice that the states of energy $\epsilon_{n,+}$ have a zero probability of occupation for all time.
    Indeed, since the dynamics only originate from the orbital degree of freedom, and no valley orbit coupling exist in this ideal model, no transition to a state of valley index $v=+$ can happen because we started in a state with $v=-$.

    A second striking feature is the oscillatory behaviour of the probabilities.
    Noticeably, the period of the oscillations is constant (it depends only on the potential of the quantum dot), but the amplitude can be increased through $\vqd$.
    An analytical derivation of this fact is performed in the next subsection under the adiabatic approximation.
    Note, however, that higher values of $\vqd$ slowly increase the probabilities of higher orbital states (i.e states with $n\geqslant 1$), as can be seen in the last row.
    For low values of $\vqd$, the coherent oscillations between the two states are effectively Rabi oscillations encountered in qubits, but this does not hold at higher quantum dot speeds.

    \begin{figure}[h]
        \centering
        \includegraphics[width=\columnwidth]{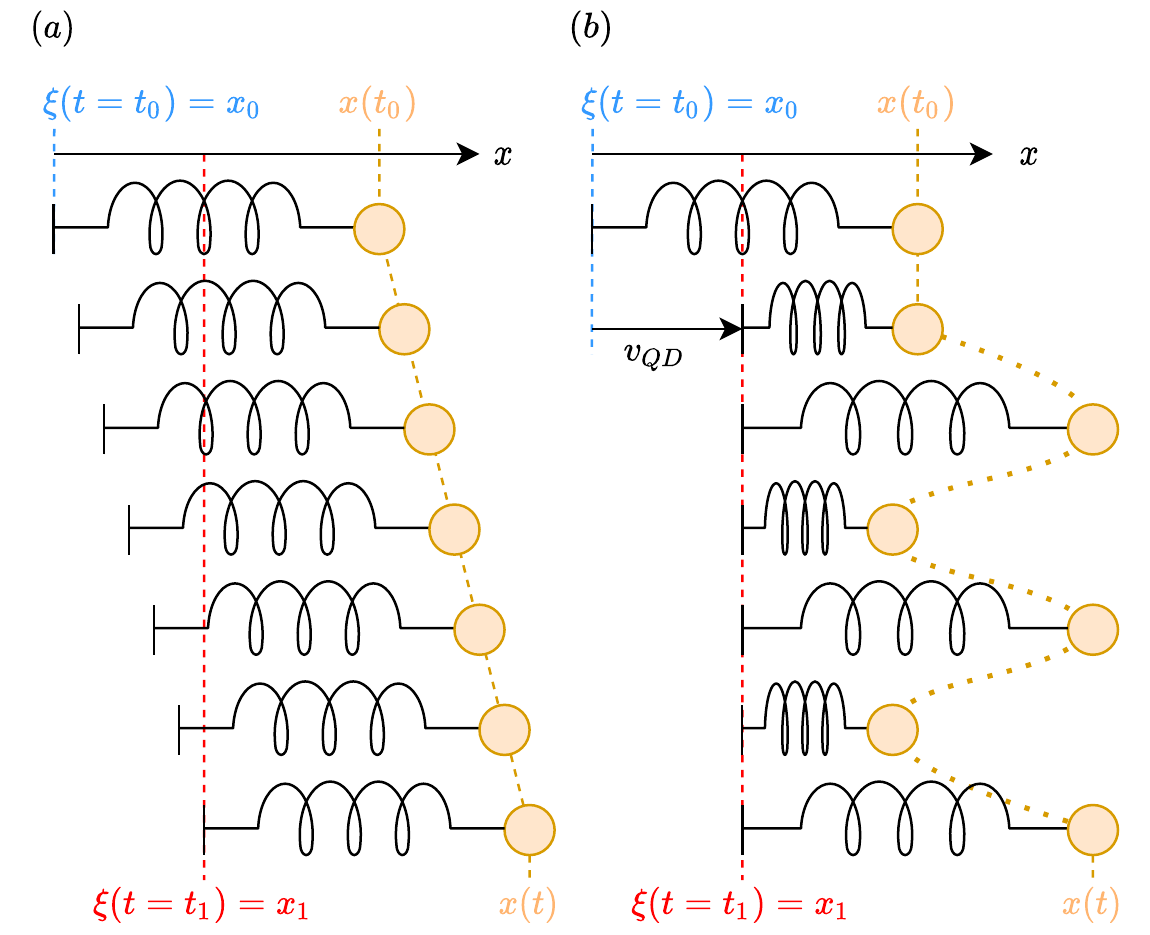}
        \caption{
            Adiabatic and non-adiabatic driving of a classical harmonic oscillator.
            a) In the adiabatic case, the mass follows the position of the spring.
            b)  In the non-adiabatic case, the spring is moved so quickly that the mass initially stays at the initial position - it does not have enough time to react to the external perturbation.}
        \label{fig:classical-analogy}
    \end{figure}
    \textbf{Classical analogy and spatial oscillations of the probability density. }
    A consequence of the presence of states other than the ground state manifests itself in the spatial oscillations of the probability density.
    The spatial oscillations of the electron can easily be understood classically as follows.
    Applying the Heisenberg equation of motion to the operators $\hat{x}$ and $\hat{p}_x$, we obtain:
    \begin{align*}
        \dv{\expval{\hat{p}_x}}{t}
        = & \expval{\frac{1}{i\hbar} \left[\hat{p}_x, \hat{H}_x\right]} \\
        = & m \omega^2 \left(\xi(t)-\expval{\hat{x}}(t)\right)\\
        \dv{\expval{\hat{x}}}{t}
        = & \frac{\expval{\hat{p}_x}}{m}
    \end{align*}
    From which we can derive the following equation of motion for $\expval{\hat{x}}$:
    \begin{equation}
        \dv[2]{\expval{\hat{x}}}{t} + \omega^2 \expval{\hat{x}} = \omega^2 \xi(t)
    \end{equation}
    This is exactly the equation of motion of a classical harmonic oscillator driven by some force $\xi$.
    Hence we see that the average position behaves exactly as in the classical case of a driven harmonic oscillator.
    The period of the spatial oscillations depends only on $\omega$ and not on the initial conditions, whilst the amplitude increases with an increase in the driving force.

    Classically, the movement of the quantum dot can be understood by pushing a spring to which a mass is attached, as shown in \cref{fig:classical-analogy}.
    Pushing the spring adiabatically, the mass is expected to follow the movement of the spring, i.e adapt to the perturbation at each time.
    However, pushing the spring over the same distance during a shorter period of time, one expects the mass to oscillate, as it does not have time to adjust to the perturbation.

    \textbf{Spatial oscillations from a quantum superposition. }
    Returning to the quantum case, one can assume that the effect of a non-adiabatic perturbation is to put the wave function into a superposition of states.
    For instance, assuming that at the end of the quantum dot's trajectory, i.e for $t=t_1$, and for relatively low values of $\vqd$, the state is of the form:
    \begin{equation*}
        \ket{\psi(t=t_1)} = \sqrt{1-p} \ket{\varphi_0} + \sqrt{p} \ket{\varphi_1}
    \end{equation*}
    where $\varphi_0$ and $\varphi_1$ are respectively the ground and first excited state of the harmonic oscillator at $t_1$, and $p$ if the state probability.
    Since for $t>t_1$ the Hamiltonian becomes time-independent (it corresponds to a harmonic oscillator centred at $\xi(t_1)=x_1$), the time evolution of the state is of the form:
    \begin{equation*}
        \ket{\psi(t)} = \sqrt{1-p} \ket{\varphi_0} + e^{-i2\pi \frac{t}{\tau}} \sqrt{p} \ket{\varphi_1}
        \label{eq:1d-superposition}
    \end{equation*}
    with $\tau = \frac{2\pi \hbar}{\Delta \epsilon}$ and period corresponding to a pulsation $\omega = \frac{\Delta \epsilon}{\hbar}$.
    Note that here the superposition is time-dependent but the state probability is time-independent, since the probability $p$ does not depend on time.

    The average position for $t>t_1$ can then be expressed as:
    \begin{align*}
        \expval{x}(t) &=  \mel{\psi(t)}{\hat{x}}{\psi(t)} \\
        &= (1-p) \mel{\varphi_0}{\hat{x}}{\varphi_0} + p\mel{\varphi_1}{\hat{x}}{\varphi_1} \\
        \quad & + 2\sqrt{p(1-p)}\cos\left(2\pi \frac{t}{\tau}\right) \Re{\mel{\varphi_0}{\hat{x}}{\varphi_1}}\\
        &= x_1 + \sqrt{\frac{2 \hbar p(1-p)}{m \omega}}\cos\left(2\pi \frac{t}{\tau}\right)
        \label{eq:avg-pos-rabi}
    \end{align*}
    where we have used equation (4.174) of reference \cite{2000BransdenJoachain} to write $\mel{\varphi_0}{\hat{x}}{\varphi_1} = \sqrt{\frac{\hbar}{2m \omega}}$.
    Hence, we recover the fact that the probability density oscillates in space, due to a quantum superposition of states.
    These spatial oscillations of the probability density are at the root of the oscillations experimentally observed by \citeauthor{2019Yamahata}.

    \subsection{\label{subsec:analytical-higher-orbital}Transition probabilities in the weak non-adiabatic regime. }
    In this section, we follow a method by \citeauthor{2000BransdenJoachain} \cite{2000BransdenJoachain} to derive an analytical expression for the probability of the first orbital excited state, i.e the state $\psi_{1,-}$.
    To simplify notations, we will drop the time-dependence of $\xi$ and denote by $m=m_x$ the electron's transverse effective mass and $\hat{H}(t) = H_x$ the Hamiltonian.
    We will also make use of the single index notation defined in section \ref{sec:models}, such that we study the transition $\psi_{0,-}\equiv \psi_0 \to \psi_2 \equiv \psi_{1,-}$.
    One can expand the Hamiltonian of the 1D model as follows:
    \begin{equation*}
        \hat{H}(t) = \frac{\hat{p_x}^2}{2 m} + \frac{1}{2} m \omega^2\hat{x}^2 + \frac{1}{2} m \omega^2\xi^2 -  m \omega^2\xi \hat{x}
    \end{equation*}
    The derivative with respect to time comes out as:
    \begin{align}
        \pdv{\hat{H}}{t} = m \omega^2 \dot{\xi}(\xi - \hat{x})^2 = - m \omega^2 \dot{\xi} (\hat{x} - \xi)^2
        \label{eq:time-dev-hamiltonian}
    \end{align}
    Following  \citeauthor{2000BransdenJoachain}, the probability amplitude of the first orbital excited state $\psi_2$ under the adiabatic approximation is (equation 9.100):
    \begin{equation}
        c_2(t) = \hbar^{-1} \int_{t_0}^t \dd{t'}
        \frac{\mel{\varphi_1(t')}{\pdv{\hat{H}}{t'}}{\varphi_0(t')}}{\omega}
        \exp\left(i \int_{t_0}^{t'} \omega \dd{u} \right)
        \label{eq:c1-definition}
    \end{equation}
    with $\{\varphi_n(t) \}_{n \in \mathbb{N}}$ denoting the eigenstates of the harmonic oscillator centred at $\xi(t)$.
    We can simplify the numerator using \cref{eq:time-dev-hamiltonian}:
    \begin{align*}
        \mel{\varphi_1(t')}{\pdv{\hat{H}}{t'}}{\varphi_0(t')}
        &= - m \omega^2 \dot{\xi} \mel{\varphi_1(t')}{\hat{x}-\xi(t')}{\varphi_0('t)}\\
        &= - m \omega^2 \dot{\xi} \mel{\varphi_1(0)}{\hat{x}-\xi(0)}{\varphi_0(0)}
    \end{align*}
    Using the generating function for the Hermite polynomials (see equation (4.174) of reference \cite{2000BransdenJoachain}), one has the following identity:
    \begin{equation}
        \mel{\varphi_n}{\hat{x}}{\varphi_m} = \begin{cases*}
            \sqrt{\frac{\hbar}{m \omega}} \sqrt{\frac{n}{2}} \text{ if } m = n-1\\
            \sqrt{\frac{\hbar}{m \omega}} \sqrt{\frac{n+1}{2}} \text{ if } m = n+1\\
            0 \text{ else }
        \end{cases*}
        \label{eq:harmonic-oscillator-x-overlap}
    \end{equation}
    Hence:
    \begin{align*}
        \mel{\varphi_1(t')}{\pdv{\hat{H}}{t'}}{\varphi_0(t')}  & = - m \omega^2 \dot{\xi} \sqrt{\frac{\hbar}{2m \omega}}
    \end{align*}
    We can then obtain a more explicit expression for the probability amplitude of the first orbital excited state:
    \begin{align*}
        c_2(t)
        &=
        \frac{-m \omega^2}{\hbar \omega} \sqrt{\frac{\hbar}{2m \omega}}
        \int_{t_0}^t \dd{t'} \; \dot{\xi}(t') \exp \qty(i \omega (t'-t_0))\\
        &=
        -\omega \sqrt{\frac{m}{2 \hbar \omega}}
        \int_{t_0}^t \dd{t'} \; \dot{\xi}(t') \exp \qty(i \omega (t'-t_0))
    \end{align*}
    At this stage, it is useful to introduce the maximum speed of a quantum harmonic oscillator in the ground state.
    Assuming that the total energy corresponds to kinetic energy, we have $\epsilon_0 = \frac{\hbar \omega}{2} = \frac{1}{2} m v_{max}^2$, where $v_{\text{max}} \equiv \sqrt{\frac{\hbar \omega}{m}} = \sqrt{\frac{\Delta \epsilon}{m}}$ the maximal speed.
    We can then rewrite the expression for $c_2(t)$ as:
    \begin{align}
        c_2(t) = \frac{-\omega}{\sqrt{2}}
        \int_{t_0}^t \dd{t'} \frac{\dot{\xi}(t')}{v_{\text{max}}} \exp\qty(i \omega (t'-t_0))
        \label{eq:general-transition-first-orbital}
    \end{align}
    For $t_0 \leqslant t \leqslant t_1$, $\dot{\xi}(t) = \vqd$ we can obtain the transition probability from the modulus squared of $c_2(t)$:
    \begin{equation}
        P_{2}(t) = 2\left(\frac{\vqd}{v_{\text{max}}}\right)^2 \left| \sin\left(\frac{\omega}{2}(t-t_0)\right) \right|^2
        \label{eq:first-excited-prob}
    \end{equation}
    Hence, the state probabilities oscillate with an amplitude proportional to $\vqd^2$ as observed in \cref{fig:ideal-probs}.
    The period of oscillations can be made more apparent by rearranging the equation above:
    \begin{align*}
        P_{2}(t)
        &=
        2\left(\frac{\vqd}{v_{\text{max}}}\right)^2 \left| \sin\left(\frac{\omega}{2}(t-t_0)\right)^2 \right|\\
        &=
        \left(\frac{\vqd}{v_{\text{max}}}\right)^2 \left| 1 - \cos\left(2\pi\frac{(t-t_0)}{\frac{2\pi}{\omega}}\right) \right|\\
        &=
        \left(\frac{\vqd}{v_{\text{max}}}\right)^2 \left| 1 - \cos\left(2\pi\frac{(t-t_0)}{\tau}\right) \right|
    \end{align*}
    with $\tau=\frac{2\pi \hbar}{\Delta \epsilon}$ as previously defined.
    In the context of our simulations, the orbital spacing $\Delta \epsilon = 2$ meV gives $\tau=2.1$ ps, which matches the period observed in \cref{fig:ideal-probs}.

    In appendix \ref{app:higher-orbitals}, a similar calculation is performed to estimate the probability of the second excited orbital state $\psi_{2,-}=\psi_4$.

    \subsection{\label{sec:ideal:criterion}Criterion for the transition to the non-adiabatic regime}
    Strictly speaking, an adiabatic evolution implies that the probability of being in eigenstate remains the same throughout the infinitely slow perturbation.
    For instance, if the initial probability of the state $\psi_2$ is initially zero, then it should continue being zero at all future times.
    In practice, it is convenient to adopt a less restrictive definition by comparing how well an adiabatic approximation matches with the actual results.

\begin{figure}[h]
        \centering
        \includegraphics{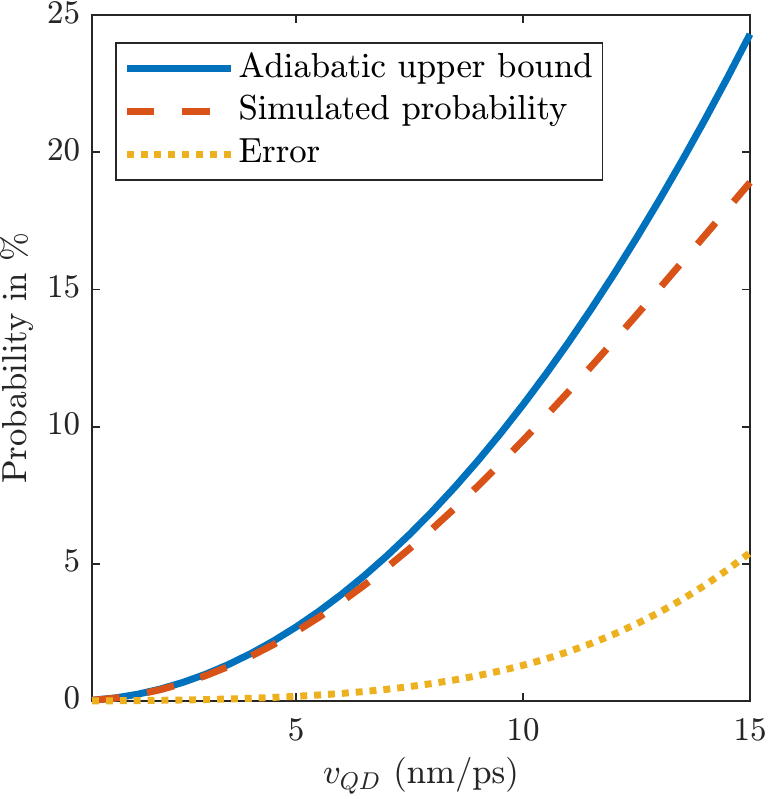}
        \caption{
            Maximum value of the probability of the first excited orbital state $\psi_{2}$ for different values of $\vqd$.
            The blue solid line is the upper bound of \cref{eq:upper-bound-ideal} derived in the adiabatic approximation.
            The red dashed line correspond to the maximum over time of the probability of observing the state $\psi_{2}$ in our numerical simulations.
            The dotted yellow line gives an indication of how well the adiabatic approximation agrees with the numerical results.
            For small values of $\vqd$, the upper bound is a very good approximation, but the error widens at $\vqd$ increases, reflecting the smooth transition to the non-adiabatic regime.
        }
        \label{fig:ideal-final-probs}
    \end{figure}

   Using \cref{eq:first-excited-prob} we can find an upper bound for the probability of the first orbital in the adiabatic approximation, namely the maximum value of $P_{0\to2}$ is given by
    \begin{equation}
        \max_t P_{0 \to 2}(t) = 2 m \frac{\vqd^2}{\Delta \epsilon} \quad .
        \label{eq:upper-bound-ideal}
    \end{equation}
    As the bound is directly proprtional to the square of $\vqd$ (the orbital spacing $\Delta \epsilon$ is fixed to 2 meV), one can appreciate that the physics is particularly sensitive to the quantum dot speed.
    This upper bound is plotted in \cref{fig:ideal-final-probs} (blue solid line).

    As discussed in the previous subsection, and considering the results of \cref{fig:ideal-probs}, the value of the final probabilities depend on the time at which we look for a fixed value of $\vqd$.
    As a consequence, we plot in \cref{fig:ideal-final-probs} the the maximum value of the probability of $ P_{0 \to 2}$ depending on $\vqd$, as obtained from the simulations (red dashed line).

    We remark that the numerical data matches the adiabatic approximation for low $\vqd$, i.e when the adiabatic approximation is valid.
    However, as $\vqd$ increases, the adiabatic approximation, utilised to derive the analytic expression, becomes invalid and the difference between the upper bound and the numerical results widens as expected.
    Indeed, this can easily be interpreted by looking at the results of \cref{fig:ideal-probs} where we see higher orbital states appearing as $\vqd$ increases.
    As a consequence, some of the probability leaks to these higher orbital states, explaining why the results diverge.
    The yellow dotted line gives an estimation of how ``non-adiabatic'' the system is. Low values testify to an adiabatic system for the orbitals, while values above a certain threshold (arbitraly chosen) indicate a non-adiabatic regime.

    To conclude this section, we would like to emphasise that the plot of \cref{fig:ideal-final-probs} only considers the orbital degrees of freedom, as the valley degrees of freedom do not play a role in the ideal model. Still, \cref{fig:ideal-final-probs} will prove useful for analysing the results involving the valley degrees of freedom of the step model, discussed in section \ref{sec:non-adiabatic-step}.

    \section{\label{sec:spectrum}Spectrum in the step model}
    We now perform the same analysis but with the step model. Critically, the perturbation now couples the orbitals with the valley states of the electron.\\

\begin{figure}[h]
        \centering
        \includegraphics[width=\columnwidth]{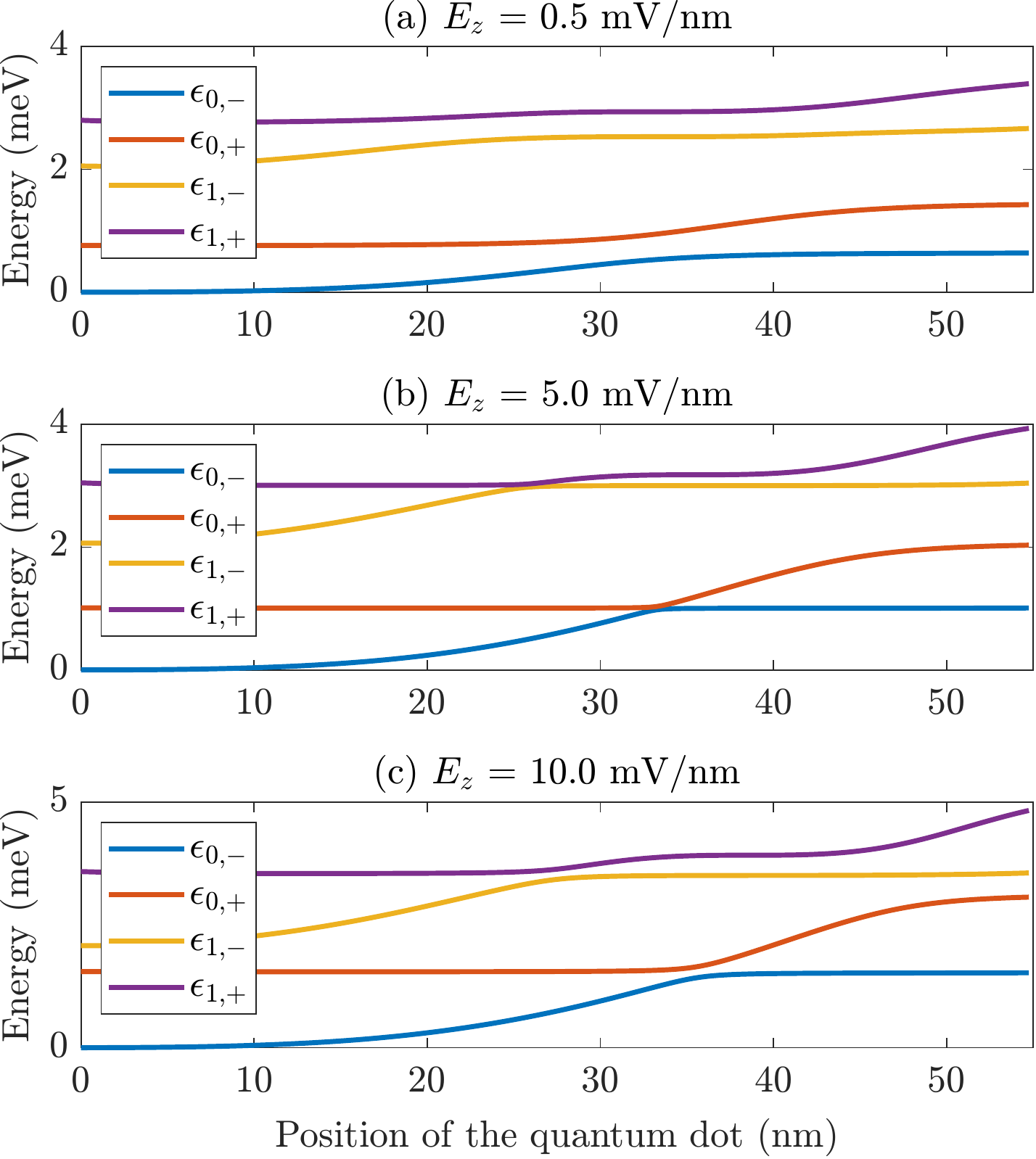}
        \caption{
            Spatial evolution of the spectra of the single step model for different electric fields $E_z$.
            In this figure, the reference energy has been set to 0 meV to increase readability.
            An anti-crossing opens after the step located at $x_s=30$ nm, and can be understood by the probability densities of the two lowest states.
            The numerical results show that the energy gap at the anti-crossing can be tuned by the electric field $E_z$ applied perpendicular to the SiGe interfaces (see \cref{fig:2d-heterostructure}).
            Note also the global increase of the energies as the electron is confined to a tighter region, as we would expect.}
        \label{fig:step-spectra}
    \end{figure}

    \textbf{General remarks on the spectrum. }
    We now discuss the salient features of the spectrum of the step model as the minimum of the quantum dot potential is varied.
    The results are plotted in \cref{fig:step-spectra} for different electric fields $E_z$.
    To simplify notations, we continue to label the energies with the indices $(n,v) \in \mathbb{N} \times \qty{-,+}$, but one should be reminded that near the step, the strong valley-orbit coupling prevents $(n,v)$ from being good quantum numbers.
    Note that far away from the steps, the wave functions are well defined valley states, and the indices $(n,v)$ are still good quantum numbers.

    \textbf{Origin of the anti-crossing. }
    A critical observation is the existence of an anti-crossing near the step location (at $x_s=30$ nm).
    As mentioned in the previous paragraph, the step causes the coupling of the valley and orbital degrees of freedom.
    In the absence of disorder at the well interface there exists globally defined valley states, where the valley states are translationally invariant, however, the step potential explicitly breaks this symmetry resulting in the observed anti-crossing.

    The physics behind this avoided crossing can be understood by examining the probability density of the two lowest energetic states, and is well described in \citeauthor{2016Boross}.
    Indeed, as shown in both atomistic simulations \cite{2004BoykinAPL,2004BoykinPRB} and effective mass theories \cite{2007Friesen,2010FriesenCoppersmith}, the wave functions exhibit modulations on the atomic scale of the envelope function.
    The two lowest $k_z$ valley states are symmetric and anti-symmetric superpositions of $\pm k_0$ with $k_0 \simeq 0.82 \frac{2\pi}{a}$, where $a=5.41$ \AA\ is the lattice spacing of the silicon crystal.
	The atomic scale oscillations of the two valley states are similar except for a spatial de-phasing.
    A consequence of this spatial de-phasing is that one of the state will have a maxima at the step location, while the other will have a minima.
    Hence, as the quantum dot goes through the step, the lowest energetic state will feel the step, increasing its energy, while the other will not.

    \textbf{Control of the anti-crossing gap with electric field. }
    As shown in \cref{fig:step-spectra}, the energy gap at the anti-crossing of the two lowest valley states can be electrically controlled through the applied electric field $E_z$.
    The dependence of the gap with the electric field is non-trivial and also depends on the quantum well width, which we have fixed to $N_z=40$ atoms in this paper.
    As discussed in later sections, the values of $E_z$ for which the energy gap of the anti-crossing is small will prove particularly useful for electrical control of the final state.
    One could question whether this behaviour is expected in real devices, or if it is an artifact of the two-band model we are using to model the valleys.
    In appendix A, NEMO 3D calculations using a 20 band $sp^3d^5s^*$ model show that the gap at the anti-crossing is indeed controllable with the electric field $E_z$.
    Akin to the results of the two-band model, the anti-crossing gap varies in the same non-trivial way, but it seems the trend can be reproduced only for certain quantum well widths. As a result, we expect the control of the anti-crossing gap to be feasible in real devices.

    \textbf{Higher orbital anti-crossing.}
    We notice that a similar anti-crossing happens between valley states of higher orbitals too (states of energies $\epsilon_{1,-}$ and $\epsilon_{1,+}$).
    However, this anti-crossing happens before that of the ground state orbital.
    This makes sense since the higher orbital states of the harmonic oscillators have a wider spatial extent over the $x$ position as the orbital number $n$ increases.
    In turn, the step is felt earlier, explaining the earlier appearance of the anti-crossing.

    \begin{figure}[h]
        \centering
        \includegraphics[width=\columnwidth]{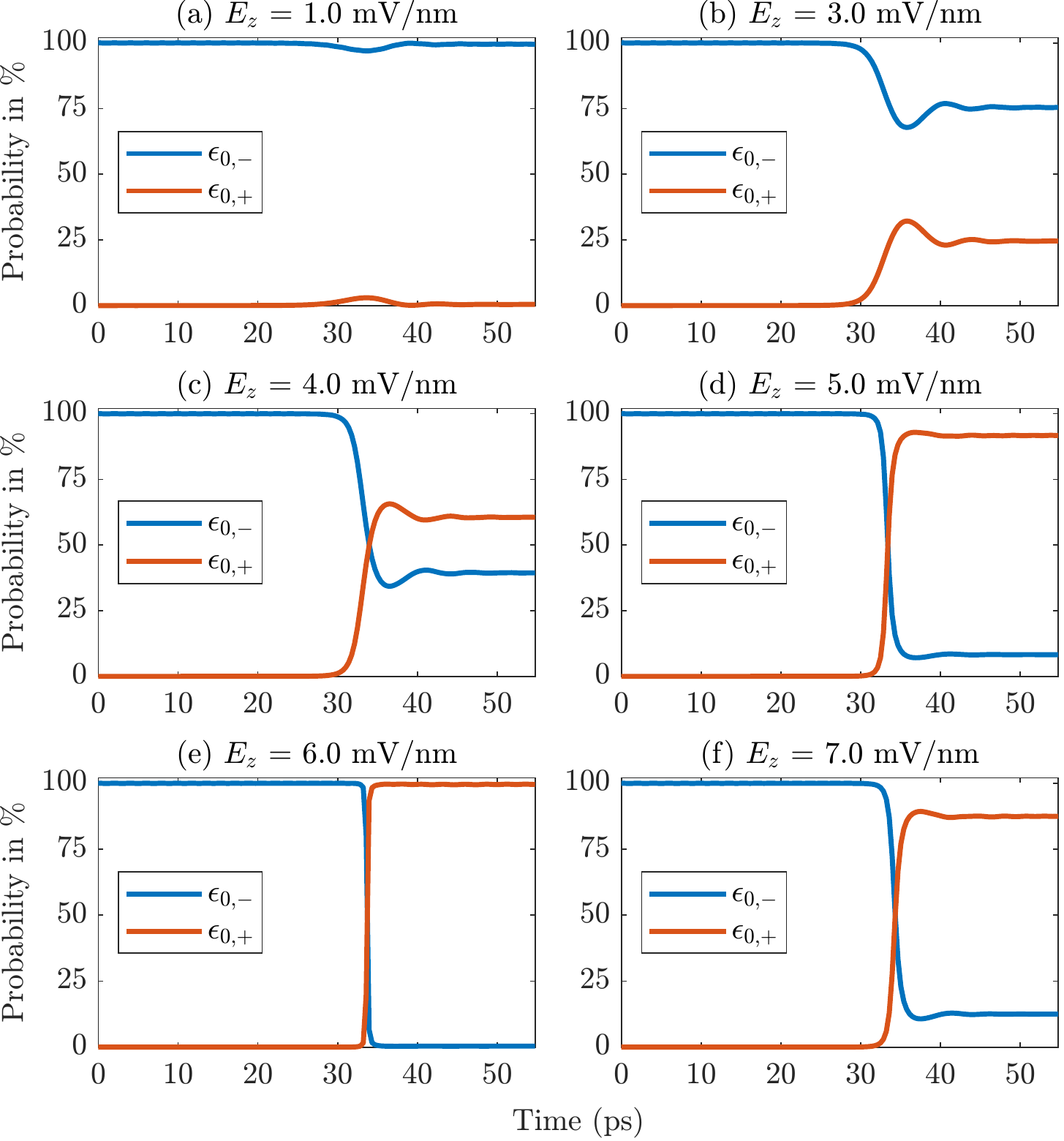}
        \caption{
            Evolution of the probabilities at $\vqd=1$ nm/ps for different electric fields $E_z$.
            The probabilities of higher energy states were at most 0.25 \%, so they have been omitted here for readability.
            To a good approximation $\vqd=1$ nm/ps results in a two-level system (see \cref{fig:ideal-final-probs}) and can form the basis states of a qubit.
            Through the applied electric field $E_z$, the energy gap of the anti-crossing can be tuned, which results in the control of the valley states.
        }
        \label{fig:step-probs-vqd-1}
    \end{figure}
    \section{\label{sec:non-adiabatic-step}Non-adiabatic effects in the step model}
    \subsection{General comments}
    \textbf{Results at a fixed low quantum dot speed.}
    The movement of the quantum dot potential has the effect of bringing energy into the system.
    This additional energy allows the ground state to overcome the anti-crossing gap observed in the spectrum.
    As intuitively expected, the smaller the gap between the two lowest eigenstates, for the same perturbation the stronger the probability of transitioning to other valley states.
    This is illustrated well in \cref{fig:step-probs-vqd-1} for $\vqd = 1$ nm/ps where the value of $\vqd$ is small enough that the system is mostly confined to the two valleys of the first orbital state (this can be verified by examining the yellow dashed line of \cref{fig:ideal-final-probs}, showing that the perturbation is adiabatic for the orbitals).
    For this fixed quantum dot speed, the variation of the applied electric field is the only parameter used to tune the final probabilities of the valley states.
    Comparing the gaps in the spectra of \cref{fig:step-spectra} and the final probabilities of \cref{fig:step-probs-vqd-1}, one can relate the final probabilities to the tuning of the anti-crossing gap.
    \begin{figure}[h]
        \centering
        \includegraphics[width=\columnwidth]{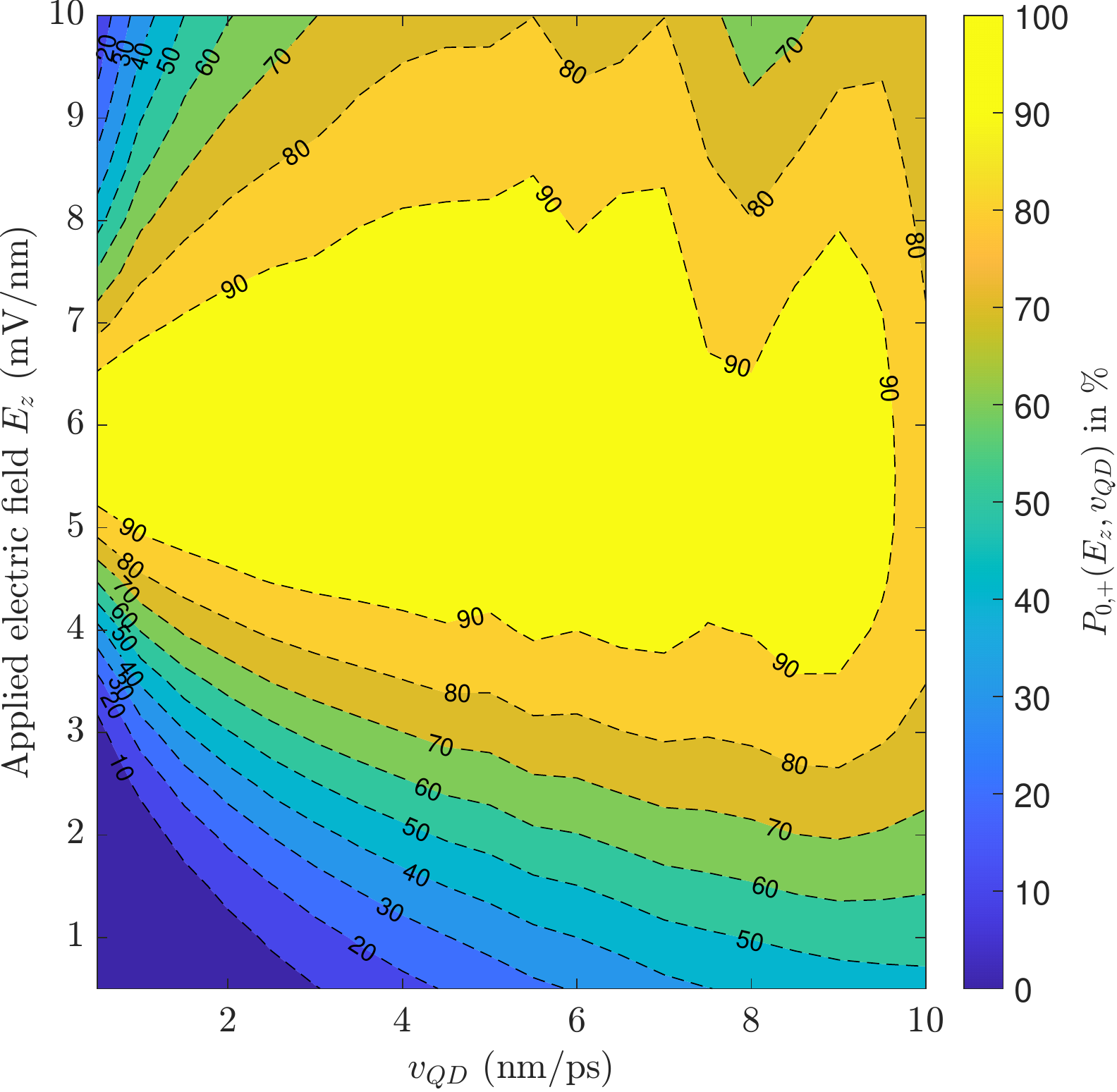}
        \caption{
            Probability of the state $\psi_{0,+}$.
            This allows to find the combination of parameters $(\vqd, E_z)$ to reach a specific final probability for $\psi_{0,+}$.
            For the quantum well width we chose, the anti-crossing gap is the smallest at values of $E_z \simeq 6$ mV/nm.
            This explains the high transition probabilities observed for this range of electric fields.
            For high values of $\vqd$, transition to higher orbital states take place, which explains the reduction of the probability occurring around $\vqd=9$ nm/ps.
        }
        \label{fig:step-prob-excited}
    \end{figure}

    \textbf{Results for different combinations of quantum dot speeds and electric fields.}
    In the ideal model, the final probabilities were tuned through $\vqd$ (see \cref{fig:ideal-final-probs}).
    In the step model too, the quantum dot speed $\vqd$ can be used to tune the final probabilities. In \cref{fig:step-prob-excited}, we plot the probability of the state $\psi_{0,+}$ for a combination of electric fields $E_z$ and quantum dot speeds $\vqd$.
    As expected from the spectra of \cref{fig:step-spectra}, the ``sweet spot'' for transitioning to the state $\psi_{0,+}$ corresponds to small anti-crossing gaps, which happens in a range of electric fields $E_z \in \qty[4, 6]$ mV/nm.
    One can also remark that increasing the quantum dot speed increases the transition probability for relatively small values of $\vqd$.
    For higher values of $\vqd$ however, the transition probability actually decreases with increasing $\vqd$, since transition to higher orbital states starts to take place instead.
    This is easily interpreted from \cref{fig:ideal-final-probs}, which testifies to the presence of higher orbital states.

\begin{figure}[h]
        \centering
        \includegraphics{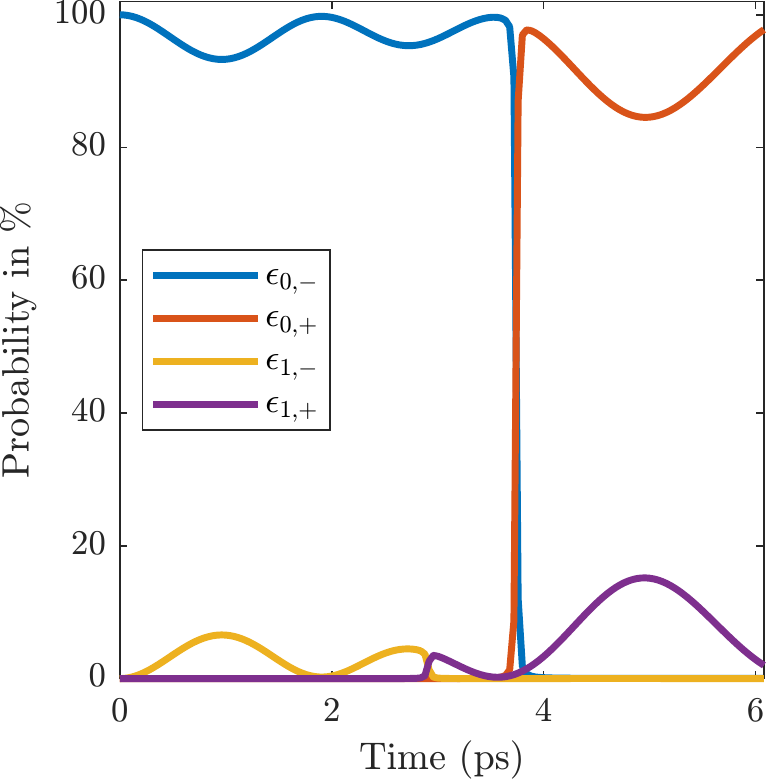}
        \caption{
            In this figure, $\vqd=9$ nm/ps and $E_z=$ 6 mV/nm.
            We can observe an exchange of valley states for the orbital ($n=1$), due to an anti-crossing in the spectrum happening earlier than that of the ground orbital ($n=0$).
            Another interesting feature is the change in both the amplitude and the period of the oscillations after the valley flipping occurs.}
        \label{fig:step:doule-flip}
    \end{figure}

    \textbf{Higher orbital valley flipping.}
    The spectra of \cref{fig:step-spectra} show that an anti-crossing exists between the states of the first excited orbital.
    Since the transition between valley states is due to the anti-crossing, it is legitimate to expect valley flipping behaviour for the first excited orbital.
    Still, according to the spectrum, this valley exchange should happen before that of the ground orbital states, as the anti-crossing happens at an earlier position as explained in section \ref{sec:spectrum}.
    This behaviour is indeed verified at higher quantum dot speeds as shown in \cref{fig:step:doule-flip}.

    \subsection{Landau-Zener approximation}
    \textbf{Standard Landau-Zener formula and anti-crossings.}
    The probability of non-adiabatic transitions in a two-level system with an avoided crossing has been heavily studied, see reference \cite{1981Rubbmark} for instance.
    In particular, the Landau-Zener formula can be used to estimate the transition probabilities due to non-adiabatic effects.
    The benefit of such an analysis is that one can simply compute the final state probabilities without solving the full time-dependent Schrödinger equation.

    Under the Landau-Zener approximation, the energy separation between the uncoupled states is a linear function of time with gradient $\alpha$, which is assumed to extend over all time.
    Furthermore, we denote the coupling induced by the step potential as $W$ and also assume it is constant in time.
    Knowing this, we can define
    \begin{equation}
        \Gamma = \frac{W^2}{\hbar |\alpha|}
    \end{equation}
    where the probability of transitioning to the excited state is given by the formula
    \begin{equation}
        P_{LZ} = e^{-2\pi \Gamma}
    \end{equation}

    \begin{figure}[h]
        \includegraphics[width=0.9\linewidth]{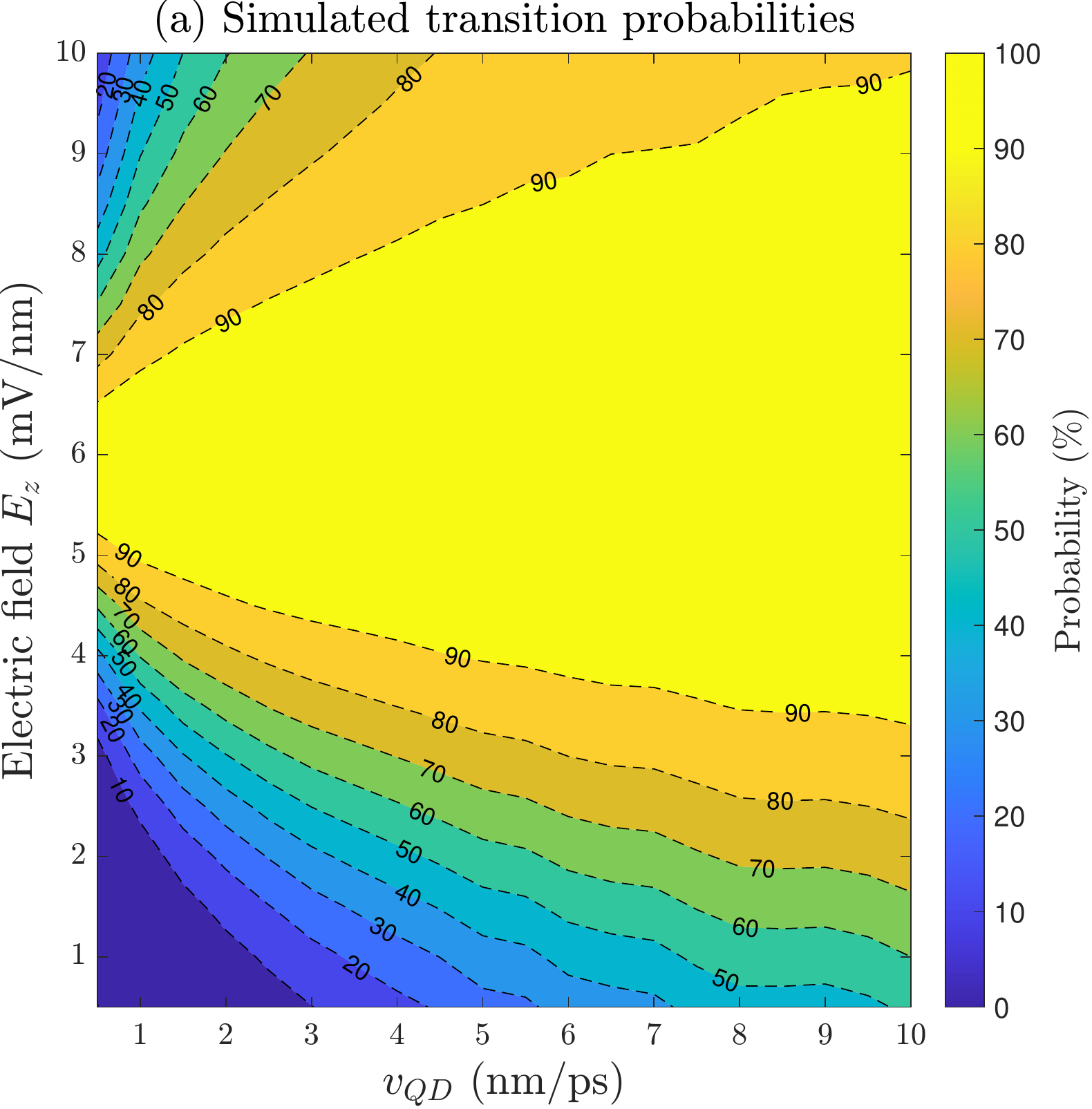}
        \includegraphics[width=0.9\linewidth]{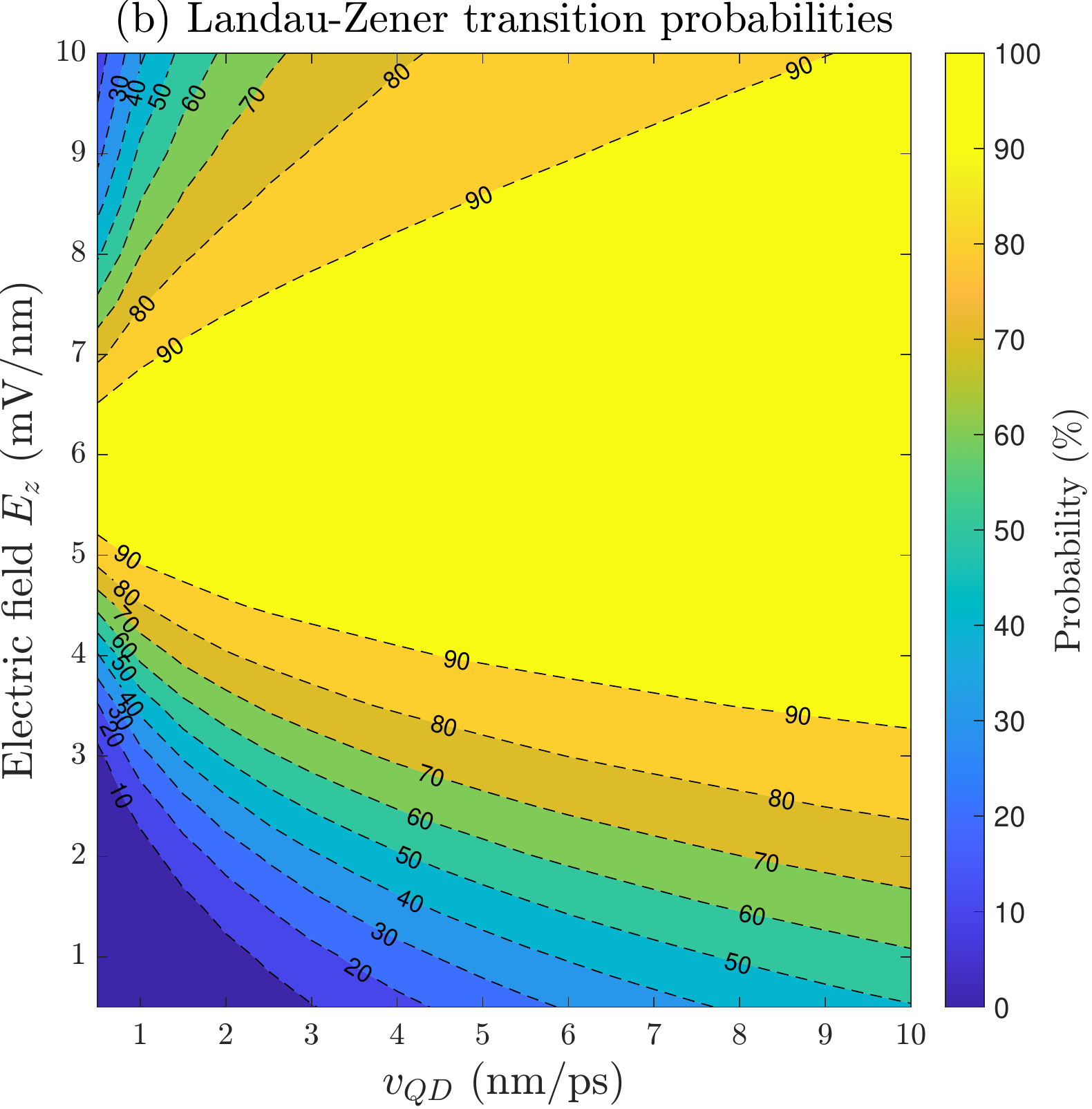}
        \caption{
            Result of the Landau-Zener approximation.
            (a) The simulated transition probabilities in the step model.
            (b) Transition probabilities under the Landau-Zener approximation in the step model. We can observe a remarkable agreement between our numerical results and those obtained from the Landau-Zener approximation.
        }
        \label{fig:landau-zener}
    \end{figure}

    \textbf{Fit to Landau-Zener.}
    In a realistic model the perturbation is only linear over a finite range of time, as our numerical results for the spectra show in \cref{fig:step-spectra}.
    An analytical modelling of our spectra is complicated, as the expression of the wave functions is not analytical in a quantum well with an applied electric field.
    As a consequence, we take the approach of fitting our numerical results to a simpler model.
    Luckily, such a simpler model is provided by \citeauthor{1981Rubbmark} \cite{1981Rubbmark}.
    Hence we fit the energy level separation between the two lowest energy states ($\epsilon_{0,+}-\epsilon_{0,-}$) to equation (18) of reference \cite{1981Rubbmark} reproduced below ($\alpha$ and $\tau$ are the fitting parameters).
    The resulting transition probabilities are plotted in \cref{fig:landau-zener}.b for different values of $\vqd$ and $E_z$.
    \begin{equation}
        \epsilon_{0,+}(t)-\epsilon_{0,-}(t) = \alpha \tau \qty(\frac{1}{1+ \exp\qty(-4t/\tau)} - \frac{1}{2})
    \end{equation}

\begin{figure}[h]
        \centering
        \includegraphics[width=\linewidth]{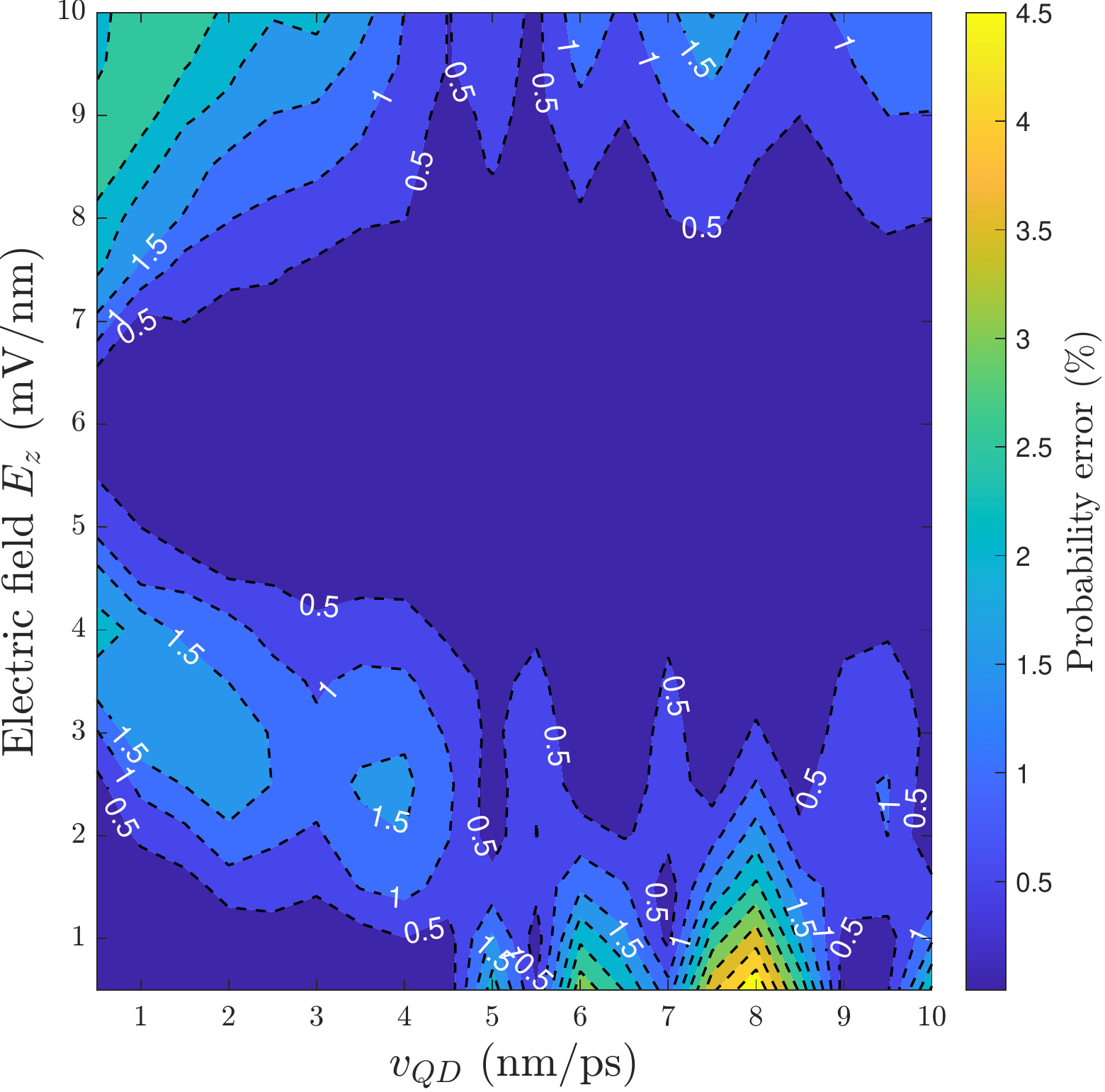}
        \caption{
            Difference between the simulated transition probabilities and those obtained under the Landau-Zener approximation.
            We can see that the Landau-Zener approximation gives good agreement with our numerical results.
            The maximum value of the relative error was 1.0 \%, not shown in this figure.
        }
        \label{fig:landau-zener:error}
    \end{figure}

    A surprising result is that the evolution of the transition probabilities is well captured by the Landau-Zener fit for almost all electric field values and the quantum dot speeds.
    In \cref{fig:landau-zener}.a, the simulated transition probabilities $1-P_{0,-}$ are plotted.
    The Landau-Zener approximation in \cref{fig:landau-zener}.b shows the same evolution as that of our numerical results.
    In \cref{fig:landau-zener:error}, we plot the difference between the simulated transition probabilities and the ones obtained from the Landau-Zener approximation $P_\text{LZ}$.
    This illustrates that the Landau-Zener approximation deviates only slightly from our numerical results.
    However, one should note that for large $\vqd$ one would certainly observe transitions to higher orbitals and valley states whilst Landau-Zener only considers the transition to the other state in the two-level system.
    Regardless, the maximum value of the relative error across all values of electric field and quantum dot speed was only 1.0\% (not shown in \cref{fig:landau-zener:error}), and hence it provides a good measure of weather the electron is in the ground state or not - with perhaps no qualification on which state is has transitioned to.

    \section{\label{sec:qubit}Qubit application}
    \textbf{Validity of the two-level system and qubit application.}
    The choice of the wide energy level spacing $\Delta \epsilon = \hbar \omega = 2$ meV for the orbital states allows one to prevent transitions to higher orbital states, thus allowing for a two level valley system to form.
    However large values of $\vqd$ can still lead to leakage to higher orbitals, and the electric field $E_z$ also plays a complex role.
    Regardless, there does exist a range of parameters for which the probability of higher orbital states is negligible and a set of qubit basis states is realised.
    We can estimate the range of parameters for which the final state is a two-level system numerically by summing the probabilities of the two states $\psi_{0,-}$ and $\psi_{0,+}$ forming our preferred two-level system.
    Using the two lowest energy states as a computational basis, our model effectively describes the time-evolution of a pure valley qubit with all-electrical control.

    \textbf{Coherence time and protection from charge noise.}
    The use of the valley degrees of freedom to encode the quantum information is interesting in that it provides certain immunity to charge noise \cite{2012Culcer}.
    Charge noise is one of the leading sources of decoherence of silicon qubits \cite{2009Culcer,2012Culcer,2018LaddCarroll,2021Chatterjee}.
    The resistance to charge noise has been experimentally verified through Laudau-Zener interferometry in a Si/SiGe double quantum dot by \citeauthor{2018MiKohlerPetta} \cite{2018MiKohlerPetta}.
    Previous implementations of qubits leveraging valley states in double quantum dots have shown a coherence time on the scale of nanoseconds \cite{2013Dzurak,2015Kim,2017SchoenfieldFreemanJiang,2019Penthorn}.
    Although those qubits were implemented in double quantum dots, whilst our system is a single quantum dot, we expect the coherence time of our architecture to have a similar order of magnitude.

    \textbf{Experimental advantage. }
    The all-electrical control of the final state probabilities is critical for scalability, as one would not need magnetic fields for qubit manipulation as in most spin qubit implementations \cite{2018LaddCarroll,2021Chatterjee}.
    The absence of applied magnetic field, and hence microwave lines, makes the integration of such a system significantly more simple.
    A second advantage is that we are not constrained by any adiabatic condition, as a consequence, provided the decoherence times are similar to that of other hybrid valley qubits, we can perform orders of magnitude more gate operations.

    \textbf{Coupling, readout and other issues. }
    Serious difficulties are to be overcome for this scheme to be usable in practice.
    First, engineering the electrical pulses used in our simulation may be challenging to implement.
    Indeed, the engineering required to control the spatial evolution of the quantum dot potential, i.e the trajectory $\xi$, may be complicated due to the time scales involved.
	As argued by \citeauthor{2016Boross}, it may be somewhat achievable using a double quantum dot architecture and varying the barrier and detuning to move the electron through the step \cite{2015Zajac,2019Yamahata}.
    This will inevitably be complicated by the additional gates needed to apply the electric field $E_z$, due to the existence of cross couplings.
    On the other hand, we have been limiting ourselves to constant speeds for $\xi$, and hence more complicated trajectories are unexplored and may be fruitful.

    Finally, both readout and coupling to other such qubits has not been explored and may pose a significant technical challenge for the scalability of the proposed scheme.
    A potential avenue towards readout could be implemented with a form of charge sensing, as is done in most silicon qubits \cite{2021Chatterjee}. This would require the incorporation of another dot after the step, and adjusting the energy levels so that only the highest energy valley states tunnels to the second dot.

    \begin{figure}[h]
        \centering
        \includegraphics[width=\columnwidth]{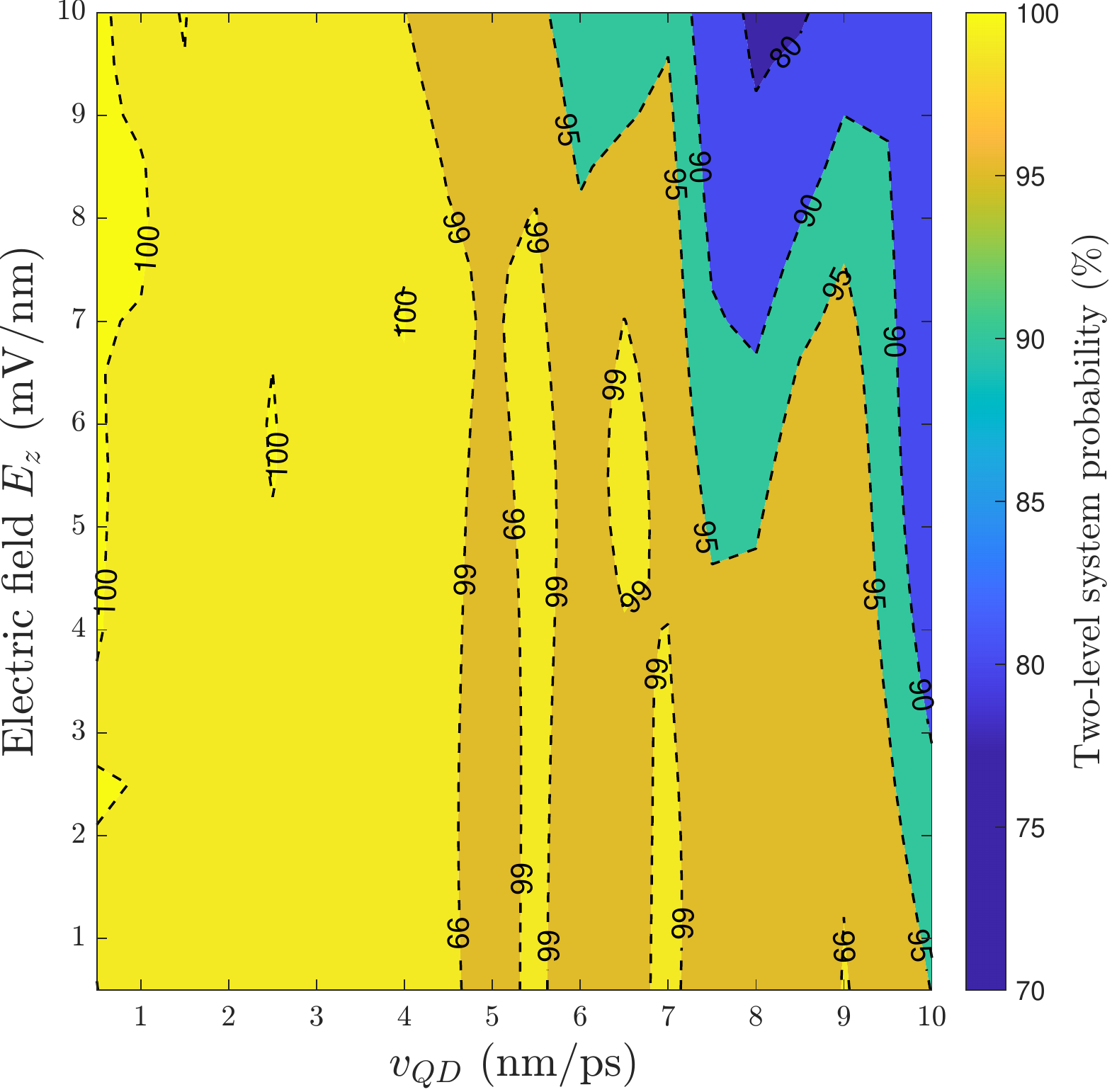}
        \caption{
            Evolution of the probability of being in a two-level system based on the two lowest valley states $\psi_{0,-}$ and $\psi_{0,+}$.
            The bright yellow/orange regions give a range of parameters for which the final state is a good two-level system suitable for qubit applications.
        }
        \label{fig:two-level-map}
    \end{figure}

    \section{\label{sec:conclusion}Conclusion}
    To summarise our work, we have shown that displacing a quantum dot potential through an interface step leads to an anti-crossing in the spectrum of the two lowest energetic states.
    A critical observation, which was verified with tight-binding calculations of NEMO 3D, is that the energy gap at the anti-crossing can be controlled by applying an electric field perpendicular to the interface.
    Since the transition probabilities depend on the energy gap, and the latter is electrically controlled, we can achieve all-electrical control of the final state probabilities.
    There is a range of parameters for which the final state behaves as a charge qubit encoded on valley states, though readout and coupling have not been explored.

    There are two main limitations of our results.
    The first is the qualitative nature due to our highly idealized modelling.
    Indeed, we enforced hardwall boundary conditions for the $z$ confinement, so that the wavefunction does not penetrate through the SiGe layers, while a more realistic choice would be adding a finite height barrier of around 150 meV.
    We also neglected both the strain and alloy disorder of the Ge atoms.
    We also note that our work is only valid in the low density limit hence ignoring both Coulomb and exchange effects.
    Despite these remarks, the results obtained from NEMO 3D indicate that the physics discussed should still be applicable to real devices, with only qualitative modifications.
    Finally, the precise engineering of both the device geometry and the applied voltages for adequate control poses additional experimental challenges.

    The second limitation is due to the experimental feasibility, namely the precise control of the fast electrical pulses to move the quantum dot and the existence of a single atomic step.
    Indeed, we expect multiple steps to form over the spatial range chosen for our simulations.
    We have also neglected relaxation mechanism, as these are supposed to occur on a timescale greater than the operation times.

    Nevertheless, non-adiabatic control of valley states may open novel quantum information processing schemes.
    In this work, we have limited ourselves to the study of a single quantum dot, but we suspect that the results could be adaptable to more complicated double dot systems making the possible implementation of these ideas more realistic.
    Additionally, in some emerging 2D and topological materials, spin and valleys are strongly momentum locked, thus fast manipulation of valleys could lead to non-adiabatic control of spins.

    \begin{acknowledgments}
        We acknowledge useful discussions with Edyta Osika concerning alternative simulation methods, and Vyacheslavs Kashcheyevs concerning the anti-crossing.
        This work is was partly funded under the Nicolas Baudin initiative and Thales Australia.
        Associate Professor Rajib Rahman acknowledges support from the US Army Research Office under grant number W911NF-17-1-0202.

        The research was undertaken with the assistance of re-sources and services from the National Computational Infrastructure (NCI) under NCMAS 2020 and 2021 allo-cation, supported by the Australian Government. This work was also supported with supercomputing resources provided by the Phoenix HPC service at the University of Adelaide.
    \end{acknowledgments}

    \appendix

    \section{\label{app:validation} Validation of the model}
    \textbf{Objective. }
    The \citeauthor{2004BoykinPRB} two-band tight-binding model has been verified in a Si/SiGe quantum well with flat hardwall boundary conditions \cite{2004BoykinPRB} but it has not yet been used to model a quantum well with a monatomic step interface.
    Most results discussed in the main text depend on two critical facts: the existence of an anti-crossing between the two lowest valley states, and the relative control of the anti-crossing gap with the electric field $E_z$.
    In this appendix, we compare the results obtained from the step model with those obtained with NEMO 3D's 20 band $sp^3d^5s^*$ model \cite{2002Oyafuso}.

    \textbf{NEMO model. }
    In our step model of subsection \ref{subsec:step-model}, we used hardwall boundary conditions, which corresponds to an infinite barrier height. To model this in NEMO 3D, we utilised a SiO2 interface which has a large barrier height of $3$ eV. To reduce simulation time, we also adopted a tighter confinement for the quantum dot. The curvature of the quantum dot was set to $k=10^{-1}$ meV/nm$^2$ in all results presented in this appendix.
    \begin{figure}[h]
        \centering
        \includegraphics[width=\columnwidth]{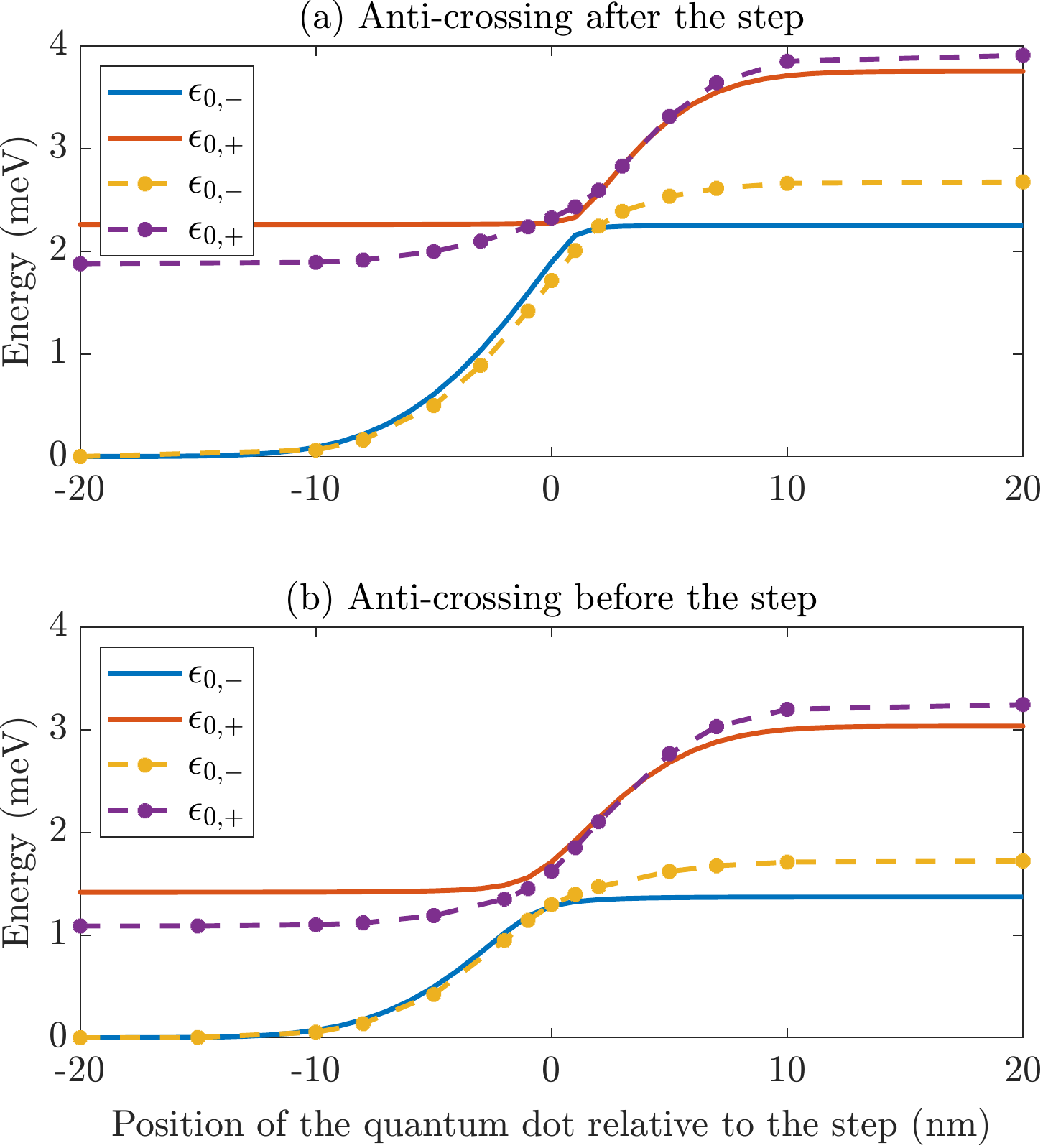}
        \caption{
            Evolution of the spectrum with the location of the quantum dot relative to the step at $0$ nm.
            The anti-crossing can happen either before (a) or after (b) the step, a fact captured in both our step model (solid lines) and NEMO 3D's 20 band $sp^3d^5s^*$ model (dotted lines).
            In (a), the quantum well width is 16.5 monolayers for the solid lines and 18 monolayers for NEMO 3D.
            In (b), the quantum well width is 21 monolayers for the solid lines, and 20 monolayers for NEMO 3D.
            The applied electric field is $E_z = 10$ mV/nm for all the curves in this figure.
        }
        \label{fig:verification:anticrossing}
    \end{figure}

    \textbf{Limitations. }
    In their paper, \citeauthor{2004BoykinPRB} showed that the two-band model they proposed correctly reproduced the trends expected for the valley splitting evolution with the width of the quantum well.
    In particular, the valley splittings have the correct order of magnitude but differ from NEMO 3D calculations.
    As such, their model is qualitative and not quantitative, and we expect to observe the same differences with our model.
    Similarly, we shall focus on only verifying that the critical trends necessary for our conclusions, namely the anti-crossing and the control of the energy gap, are mirrored by NEMO 3D.

    \textbf{Results. }
    In \cref{fig:verification:anticrossing}, we compare spectra obtained using our step model (solid lines) and those obtained from NEMO 3D calculations (dotted lines).
    The reference energy was set to 0 to facilitate the comparison of both spectra.
    We remark that in both simulations, the location of the anti-crossing can happen before or after the step depending on the quantum well width.
    In both cases, the existence of the anti-crossing is verified by NEMO 3D.

\begin{figure}[h]
        \centering
        \includegraphics[width=\columnwidth]{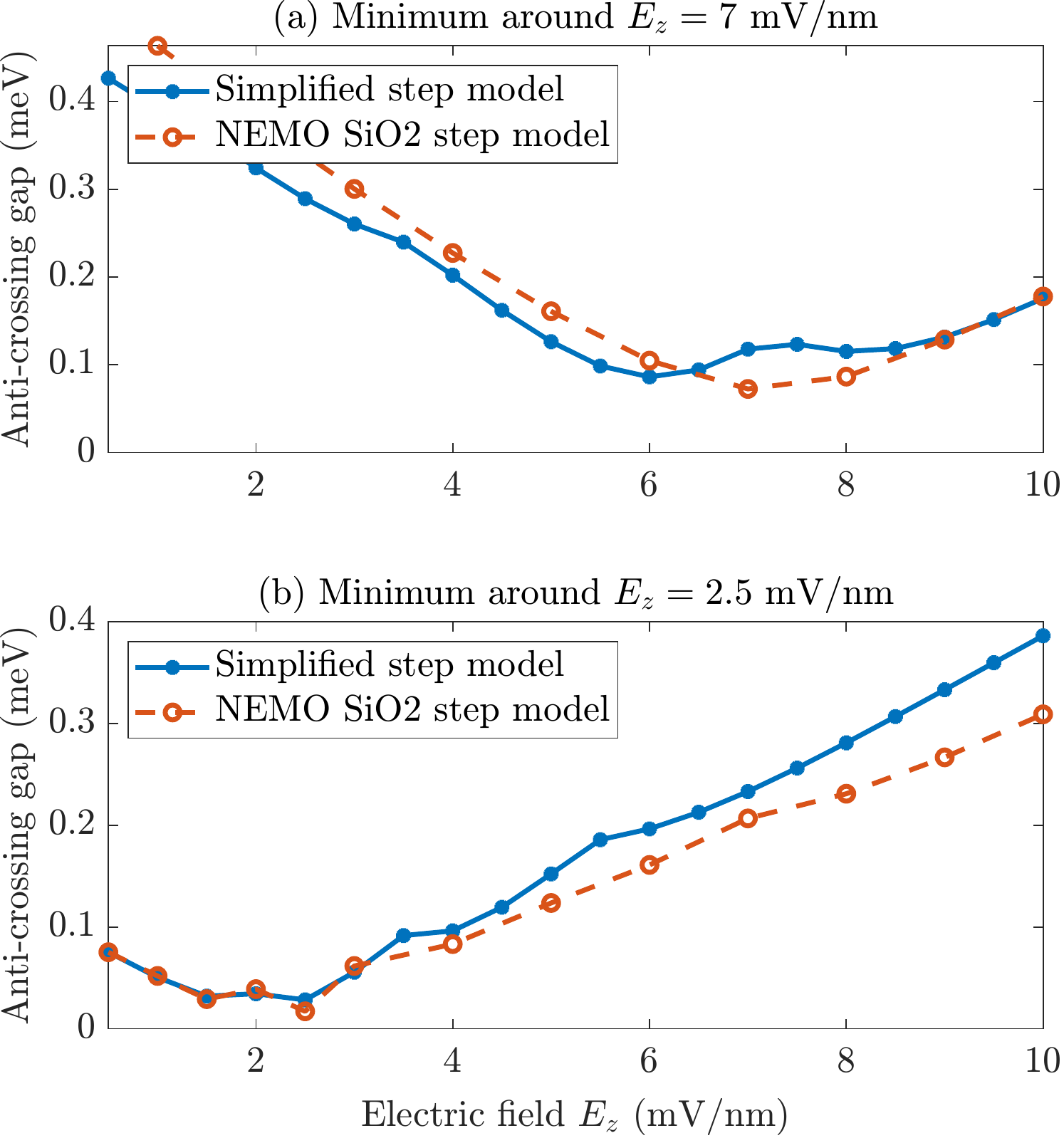}
        \caption{
            Evolution of the anti-crossing energy gap with the electric field in the step model (solid dotted blue) and NEMO simulations (dashed circled red).
            Depending on the quantum well width, the magnitude and location of the minimum of the gap can change.
            (a) In the step model, we used 16.5 monolayers, while 19 monolayers were used for NEMO 3D.
            (b) In the step model, we used 19 monolayers, while 20 monolayers were used for NEMO 3D.
        }
        \label{fig:verification:gap}
    \end{figure}

    Similarly, the tuning of the anti-crossing gap with the electric field is observed in both simulations, as shown in \cref{fig:verification:gap}.
    As mentioned in the main text, the tuning of the gap with the electric field depends on the quantum well width.
    Some key features are exhibited by both our step model and the NEMO 3D calculations.
    Namely, the range of the tuning of the gap, the existence of a minimum, and the location of this minimum all depend on the quantum well width.

    \section{\label{app:higher-orbitals}Estimation of the probability of the second excited orbital state in the ideal model}
    In this appendix we estimate the form of the probability $P_4$ of the state $\psi_4=\psi_{2,-}$ in the ideal model.
    We adopt the notations of section \ref{subsec:analytical-higher-orbital}.
    From equation (4.174) and (9.97) of reference \cite{2000BransdenJoachain}, we can estimate that the probability amplitude of the second excited state $c_4(t)$ is related to the amplitude of the first excited state $c_2(t)$ by:
    \begin{equation*}
    \dot{c_4}(t) = \frac{c_2(t)}{\hbar \omega} \mel{\varphi_2(t)}{\pdv{\hat{H}}{t}}{\varphi_1(t)}  e^{i \omega (t-t_0)}
    \end{equation*}
    Following \cref{eq:harmonic-oscillator-x-overlap}, we find that $\mel{\varphi_2(t)}{\pdv{\hat{H}}{t}}{\varphi_1(t)} = - m \omega^2 \dot{\xi} \sqrt{\frac{\hbar}{m \omega}}$.
    Using \cref{eq:general-transition-first-orbital} we find:
    \begin{equation*}
    \dot{c_2}(t) = -i \omega \left(\frac{\vqd}{v_\text{max}}\right)^2
    \left(e^{2i\omega(t-t_0)} - e^{i\omega(t-t_0)}\right)
    \end{equation*}
    The integration is straightforward and gives:
    \begin{equation*}
    c_2(t) = 2\left(\frac{\vqd}{v_\text{max}}\right)^2
    \sin \left(\frac{\omega}{2}(t-t_0)\right)^2
    e^{i\omega(t-t_0)}
    \end{equation*}

    Finally, the transition probability to the second excited state can be computed as:
    \begin{align}
    P_{0 \to 4}(t)
    &= 4\left(\frac{\vqd}{v_\text{max}}\right)^4
    \left| \sin \left(\frac{\omega}{2}(t-t_0)\right) \right|^4
    \label{eq:second-excited-prob}\\
    &= P_{0 \to 2}(t)^2 \nonumber
    \end{align}
    Comparing \cref{eq:first-excited-prob} and \cref{eq:second-excited-prob} we see that the probability of the electron being in the state $\psi_{4}$ at time $t$ is the square of it being in the state $\psi_{2}$.

    \bibliography{references}

\end{document}